\documentclass{aa}  

\usepackage{graphicx}
\usepackage[version=4]{mhchem}
\usepackage{txfonts}
\usepackage[colorlinks=true,     linkcolor=blue, citecolor=blue, filecolor=blue, urlcolor=blue]{hyperref}
\usepackage{placeins}
\newcommand{\hhco}{H$_2$CO }
\newcommand{\w}{cm$^{-1}$ }

\begin{document}

   \title{Cosmic-ray induced sputtering of interstellar formaldehyde ices}


   \author{M. Faure\inst{1}, A. Bacmann\inst{1}, A. Faure\inst{1}, E. Quirico\inst{1}, P. Boduch\inst{2}, A. Domaracka\inst{2}, H. Rothard\inst{2}}

   \institute{Univ. Grenoble Alpes, CNRS, IPAG, 38000 Grenoble, France\\
              \email{aurore.bacmann@univ-grenoble-alpes.fr}
    \and Centre de Recherche sur les Ions, les Mat\'eriaux et la Photonique, CIMAP-CIRIL-GANIL, Normandie Universit\'e, ENSICAEN, UNICAEN, CEA, CNRS, 14000 Caen, France\\
             }

   \date{Received xxx, 2024; accepted xxx, 2024}

 
  \abstract
{In the cold and dense regions of the interstellar medium (ISM), for example in prestellar cores, gas-phase chemical abundances undergo a steep decrease due to the freeze-out of molecules onto the dust grain surfaces. While the depletion of many species would bring molecular abundances to undetected levels within short timescales, non-thermal desorption mechanisms such as UV photodesorption or cosmic-ray sputtering allows the return of a fraction of the ice mantle species back to the gas phase and prevents a complete freeze-out in the densest regions. In the last decade much effort has been devoted to understanding the microphysics of desorption and quantifying molecular desorption yields. }
   {\hhco is a ubiquitous molecule in the ISM and in the gas phase of prestellar cores, and is likely present in ice mantles, but its main desorption mechanism is unknown. In this paper our aim is to  quantify the desorption efficiency of \hhco upon cosmic-ray impact in order to determine whether cosmic-ray induced  sputtering could account for the \hhco abundance observed in prestellar cores.}
   {Using a heavy-ion beam as a cosmic-ray analogue  at the Grand Acc\'el\'erateur National d'Ions Lourds (GANIL) accelerator, we irradiated pure \hhco ice films at 10\,K under high vacuum conditions and monitored the ice film evolution with infrared spectroscopy and the composition of the sputtered species in the gas phase using mass spectrometry. We derived both the effective and intact sputtering yield of pure \hhco ices. In addition, using IRAM millimetre observations, we also determined the H$_2$CO gas-phase abundance in the prestellar core L1689B.}
   {We find that \hhco easily polymerises under heavy-ion irradiation in the ice, and is also radiolysed into CO and CO$_2$. In the gas phase, the dominant sputtered species is CO and intact \hhco is only a minor species. We determine an intact sputtering yield for pure \hhco ices of $2.5\times 10^3$ molecules\,ion$^{-1}$ for an electronic stopping power of $S_e\sim2830$\,eV\,($10^{15}$ molecules cm$^{-2}$)$^{-1}$. The corresponding cosmic-ray  sputtering rate is  $\Gamma_\mathrm{CRD}=1.5\times 10^{18}\zeta$ molecules cm$^{-2}$ s$^{-1}$, where $\zeta$ is the rate of cosmic-ray ionisation of molecular hydrogen in the ISM. In the frame of a simple steady-state chemical model of freeze-out and non-thermal desorption, we find that this experimental cosmic-ray sputtering rate is too low (by an order of magnitude) 
   to account for the observed \hhco gas-phase abundance we derived in the prestellar core L1689B. We find however that this abundance can be reproduced  if we assume that H$_2$CO diluted in CO or CO$_2$ ices co-desorbs at the same sputtering rate as pure CO or pure CO$_2$ ices.}
  {}


    \titlerunning{Cosmic-ray sputtering of formaldehyde ices}
    \authorrunning{M. Faure et al.}
   \maketitle
%

\section{Introduction}

\begin{nolinenumbers}
The chemistry of prestellar cores is characterised by high amounts of molecular depletion due to the freeze-out of gas-phase molecules onto interstellar dust grains. At the high densities prevailing in prestellar cores, molecules from the gas phase collide and stick to the grains, building up an ice mantle. Millimetre wave observations of rotational transitions indeed show a molecular abundance decrease for most species towards the core centres \citep[e.g.][]{Willacy1998,Caselli1999,Bacmann2002} where the densities are highest. Additionally, infrared observations have established the presence of several species in the solid phase of the dense interstellar medium \citep{1983Natur.303..218W,2015ARA&A..53..541B}, the details of which are now being   revealed by the {\it James Webb} Space Telescope  \citep[JWST;][]{yang2022,2023NatAs...7..431M,2024A&A...681A...9R}. 

The accretion timescale for CO molecules at a temperature of 10\,K, expressed in years, can be estimated as $t_\mathrm{CO} = 5\times 10^9/n_\mathrm{H_2}$ where $n_\mathrm{H_2}$ is the molecular H$_2$ density \citep{2007ARA&A..45..339B}. At the densities prevailing in prestellar cores, around $10^5-10^6$\,cm$^{-3}$, this timescale is much shorter than typical core lifetimes of around $10^6$ years \citep{2015A&A...584A..91K}. In the absence of any efficient mechanism returning frozen-out molecules back to the gas-phase, CO and other molecules should be absent from the gas phase, contrary to what is observed in these regions: within $10^6$ years, for a density of $5\times 10^5$\,cm$^{-3}$, the molecular abundance should drop by several dozen orders of magnitude, while observations show a much more moderate drop of a few orders of magnitude \citep{2012A&A...548L...4P}. Therefore, desorption mechanisms allowing the frozen-out molecules to return to the gas phase have to be considered \citep{1983A&A...123..271L}.

At typical prestellar core temperatures, around 10\,K, the available thermal energy is  insufficient to enable molecular desorption, but several non-thermal mechanisms have been invoked to account for the observed molecular abundances in the gas phase: UV photodesorption, cosmic-ray sputtering, desorption by cosmic-ray induced UV photons (secondary UV radiation), IR photodesorption, or reactive desorption, in which the exothermicity of grain surface reactions contributes to the release of the reaction's products in the gas phase. 
In the last decades, experimental studies have endeavoured to characterise the efficiency of the various desorption processes. Photodesorption rates have been measured under broadband UV irradiation for molecules such as CO, N$_2$, CO$_2$, and H$_2$O by \citet{2007ApJ...662L..23O,2009ApJ...693.1209O,2009A&A...496..281O} or \citet{2010A&A...522A.108M}. Further work using monochromatic synchrotron UV radiation provided insights into the mechanisms governing UV photodesorption \citep[e.g.][]{2011ApJ...739L..36F,2012PCCP...14.9929B,2013A&A...556A.122F}. These experiments coupled with quadrupole mass  spectrometry to detect ejected gas-phase species have revealed that some molecules can be fragmented upon UV irradiation so that the photodesorption yield of the intact molecules  may represent only a minor component of the total photodesorbed fragments, as is the case for CH$_3$OH (methanol), HCOOH (formic acid), CH$_3$OCHO (methyl formate) \citep{2016ApJ...817L..12B,2023FaDi..245..488B} and H$_2$CO (formaldehyde) \citep{Feraud19}. As an illustration, the photodesorption yield measured by \citet{2016ApJ...817L..12B} for intact CH$_3$OH is two orders of magnitude smaller than previously derived from experiments using solely infrared spectroscopy ice measurements \citep{2009A&A...504..891O}, showing that it is therefore essential to monitor the products of the desorption process and not only the loss of the molecule in the solid state. Recent experiments of resonant infrared irradiation of CO and CH$_3$OH ices suggest that IR photons could also induce desorption with a flux integrated efficiency similar to that of UV photons \citep[and references therein]{santos2023}.

Chemical desorption, or reactive desorption,  occurs when the excess energy of a grain surface reaction (typically involving two radicals) can be converted into kinetic energy perpendicular to the surface overcoming the binding energy of the product with the grain \citep{1975ApJ...195...81A}. However, the efficiency of this process is difficult to evaluate in situ. Recent experiments have shown that while this mechanism can be efficient in some cases, the nature of the substrate (whether composed of bare graphite or covered with water ice) and its porosity \citep{2013NatSR...3E1338D}, as well as the number of degrees of freedom of the reaction product plays a major role. In particular, \citet{2016A&A...585A..24M} did not observe chemical desorption of CH$_3$OH formed from the reaction CH$_3$O + H, and estimated that the efficiency of chemical desorption from water ice is at most a few percent. Therefore, in the dense interstellar medium, where the composition of ice mantles is dominated by water ice, chemical desorption is expected to have a low efficiency, for large molecules at least. 

Cosmic rays are also believed to contribute to molecular desorption \citep[e.g.][]{1985A&A...144..147L,1993MNRAS.260..635W,1993MNRAS.261...83H,Shen2004} by sputtering ice species to the gas phase. The cosmic-ray induced desorption process consists of two main contributions: the `whole-grain' heating and the `thermal spike' sputtering (erosion) \citep{Bringa04}. 
While many studies so far have relied on the formalism proposed by \cite{1993MNRAS.261...83H} for whole-grain heating, cosmic-ray induced (thermal-spike) sputtering can be investigated experimentally by using swift heavy ions as cosmic-ray analogues. Studies of ice film irradiation by heavy ions carried out by \citet{seperuelo09,seperuelo10} have determined sputtering yields for CO$_2$ and CO ices by modelling the disappearance of the ice constituents from the solid phase. Of interest to account for the observed gas-phase abundances in the interstellar medium, however, is the amount of molecules that return to the solid phase intact per incident ion. Recent heavy-ion ice irradiation experiments coupled with a quadrupole mass spectrometer (QMS) 
to monitor the sputtered species in the gas phase  \citep{Dartois19} have measured this parameter, both for pure ices (e.g. CO$_2$, CH$_3$OH, CH$_3$OOCH$_3$) and for ice mixtures (CH$_3$OH/CO$_2$, CH$_3$OH/H$_2$O, CH$_3$OOCH$_3$/H$_2$O). For the species in these studies, a large fraction of the molecules are desorbed intact, 
and the sputtering yield of a minor species in an ice mixture is close to that of the species making up the bulk of the ice \citep{Dartois19}.

While the above-mentioned experiments  provide us with essential quantitative data regarding the desorption processes, they have been included in few chemical models so far, and it is still unclear whether they can account for observed molecular abundances \citep{Wakelam21}. Moreover, a large body of experimental data is still missing, even regarding major species in the dense interstellar medium. 

Formaldehyde (H$_2$CO) is a ubiquitous species in molecular clouds and prestellar cores \citep[e.g.][]{1973ApJ...183..449D,1980A&A....89..187S,1987ApJ...318..392M}, with gas-phase column densities close to that of methanol, around $10^{13}-10^{14}$\,cm$^{-2}$ in prestellar cores \citep{Bacmann2003,2016A&A...587A.130B} and in protostellar envelopes \citep{Parise2006}. In ice mantles, H$_2$CO has only been tentatively detected towards protostellar envelopes and its abundance is estimated at around 5\%, similar to CH$_3$OH \citep{2015ARA&A..53..541B,2024A&A...681A...9R}. Formaldehyde also represents an important intermediate step towards more complex molecules on grain surfaces \citep{chuang2016}, or towards complex organic molecule precursors like HCO in the gas phase \citep{2016A&A...587A.130B}. In prestellar cores, it is believed to form chiefly via hydrogenation of CO on grain surfaces \citep{watanabe2004,2016A&A...585A..24M}, as an intermediate step to methanol formation. Gas phase formation mechanisms have also been proposed \citep{Guzman2011,RamalOlmedo2021}, but distinguishing both formation mechanisms requires  detailed knowledge of the physical processes (e.g. desorption), which is lacking. 

In this paper we present a study of swift ion irradiation of pure \hhco\ ices and determine the sputtering yield of the intact molecule. We calculate the sputtering rate over the cosmic-ray spectrum (from $100$~eV/amu to $10$~GeV/amu) and use it in a simple model to compare the derived  \hhco gas-phase abundance with both \hhco observations in a prestellar source and the abundances predicted with other non-thermal desorption mechanisms.


The paper is organised as follows:  In Sect. \ref{section_experimental} we describe the irradiation experiments of thin H$_2$CO films carried out at the Grand Acc\'el\'erateur National d'Ions Lourds (GANIL, Caen, France)  and in Sect. \ref{section:observations} the IRAM H$_2$CO observations towards a prestellar core. In Sect. \ref{section:results} we present the analysis of both the observational and the experimental data and determine the intact sputtering yield of the molecule. In Sect. \ref{section:modelling} we discuss the astrophysical implications of our study and conclude in Sect. \ref{section:conclusions}. 


\section{Experiments}\label{section_experimental}

Pure \hhco ice irradiation experiments were carried out at the GANIL ion accelerator on the IRRSUD beamline\footnote{https://www.ganil-spiral2.eu} with the Irradiation de GLaces d'Int\'er\^et AStrophysique (IGLIAS) setup \citep[see][for a detailed description]{auge_iglias_2018} between May 14 and 16, 2021. The vacuum chamber was cryocooled down to a temperature of about 10\,K, and held at a pressure of $\sim 5\times 10^{-10}$ mbar. Infrared spectra were collected with a Bruker Vertex 70v Fourier transform infrared spectrometer (FTIR), operating with a HgCdTe-detector and a 1 \w spectral resolution. The angle between the spectrometer beam and the sample surface was 12$^\circ$. Thin H$_2$CO films with thicknesses of $\sim 0.53\,\mu$m were condensed on a ZnSe window at normal incidence with a needle placed at a distance of around 15 mm. 
The pressure during injection was typically $10^{-7}$ mbar. The deposition rate was adjusted to get a good optical quality, and the sample thickness was estimated using the band strengths of \citet{Bouilloud15} and the interference fringes from the infrared spectra. Gas-phase \hhco was produced from the thermal decomposition of paraformaldehyde (Sigma Aldrich, purity 95\%), heated at 100$^\circ$C in a homemade oven maintained under secondary vacuum ($\sim$ 10$^{-7}$ mbar). The gaseous species sputtered in the chamber were detected with a MKS Microvision 2 
QMS. The branching ratios of the species fragments were evaluated during their injection.

Irradiations were conducted with a $^{86}$Kr$^{18+}$ beam of 0.86 MeV/u (74~MeV) corresponding to an electronic stopping power of 2830 eV\,(10$^{15}$ molecules\,cm$^{-2}$)$^{-1}$ for an \hhco ice of 0.81\,g\,cm$^{-3}$ \citep{Bouilloud15}, as computed with the \texttt{SRIM-2013} code\footnote{http://www.srim.org.} \citep{Ziegler10}. The flux was $5 \times 10^8$ ions\,cm$^{-2}$\,s$^{-1}$, and the fluence reached at the end of the experiment was $4.36\times 10^{12}$\,ions\,cm$^{-2}$. The corresponding doses are computed and compared to typical interstellar doses in Appendix\,\ref{appendix:dose}. Infrared spectra monitoring the ice composition were taken every two minutes starting from the beginning of the irradiation. The QMS continously scanned masses from 0-80\,u, with a scan duration of $\sim 15$\,s. 

Additionally, a control experiment under the same conditions was carried out in which we irradiated a $\sim0.81\,\mu$m CO$_2$ (Air Liquide, purity $\ge 99.998\%$) ice film with the same beam, corresponding to an electronic stopping power of 3290 eV\,(10$^{15}$ molecules\,cm$^{-2}$)$^{-1}$, also computed with \texttt{SRIM-2013}, for a CO$_2$ density of 1.0\,g\,cm$^{3}$ \citep{Satorre08}. For this experiment, the flux varied between $1.1\times 10^9$ and $2.1\times 10^9$ ions\,cm$^{-2}$\,s$^{-1}$ throughout the irradiation and the final fluence was 10$^{13}$ ions\,cm$^{-2}$.

The experimental data can be downloaded from https://zenodo.org/records/13383307.



\section{Observations}\label{section:observations}

Observations of one prestellar core, L1689B, were carried out between 2013 and 2019 during several campaigns, at the IRAM 30\,m telescope, located at Pico Veleta, Spain. L1689B is situated in the Rho Ophiuchi star-forming region, and is characterised by low central temperatures and high central densities \citep{2016A&A...593A...6S}. It shows signs of gravitational contraction and appears dynamically evolved \citep{LMT01,LEST03,LeeMyers2011} despite moderate molecular depletion and deuteration \citep{Bacmann2003}. 
A wealth of oxygen-bearing molecules, including formaldehyde and more complex organics, have been detected in this source \citep{Bacmann2003,Bacmann2012}.  The observed coordinates are RA: $16^h 34^m 48.3^s$ Dec:$-24^\circ 38^\prime 04^{\prime\prime}$, corresponding to the peak of the millimetre continuum emission in the source. Several frequency setups were used to target rare isotopes of formaldehyde lines at 2\,mm and 1\,mm with the Eight MIxers Receivers (EMIR) E1 and E2, respectively. We focused on \hhco\ rotational transitions with upper energies typically lower than  $\sim 40$\,K, as levels of higher energies are unlikely to be sufficiently populated at the low temperatures prevailing in the source (typically 10\,K). Because the low-lying transitions of the main isotopologue are optically thick and self-absorbed towards the core centre \citep{Bacmann2003}, we present here observations of the optically thinner isotopologues H$_2^{13}$CO and H$_2$C$^{18}$O, in both ortho and para nuclear-spin symmetries. The receivers were connected to the Fourier transform spectrometer (FTS)  at a spectral resolution of $\sim 50$\,kHz over a bandwidth of 2\,GHz, leading to velocity resolutions of $\sim 0.10$\,km s$^{-1}$ at 2\,mm and $\sim 0.07$\,km s$^{-1}$ at 1\,mm. The data were taken using the frequency switching procedure with a frequency throw of 7.5\,MHz to reduce standing waves. Pointing and focus were typically checked every $1.5-2$ hours on nearby quasars or planets, with additional focus checks carried out shortly after sunset or sunrise. Pointing was found to be accurate to $3-4\arcsec$. Calibrations were performed every 10 to 15 minutes, depending on atmospheric stability. The beam sizes vary from 18$\arcsec$ at 134 GHz to 11$\arcsec$ at 220\,GHz.  The beam efficiencies were taken following the Ruze formula $B_\mathrm{eff}\left(\lambda \right)=B_\mathrm{eff,0}\,\,e^{-\left({\frac {4\pi \sigma }{\lambda }}\right)^{2}}$, with $\lambda$ the wavelength, $\sigma=66\,\mu$m the surface $rms$ of the 30\,m antenna, and $B_\mathrm{eff,0}=0.863$. The forward efficiencies were 93\% and 91\% at 2\,mm and 1\,mm, respectively. The spectra were converted from the antenna temperature scale $T_a^*$ to the main beam temperature scale $T_\mathrm{mb}$ using the values of the efficiencies given above.

The observational data were reduced with the Gildas/IRAM software.\footnote{https://www.iram.fr/IRAMFR/GILDAS} The data reduction consisted in folding the spectra to recover the signal from the frequency switching procedure, except for the newer data for which folded spectra are delivered by the system. Then a low-order (typically 3) polynomial baseline was fitted to the spectra over line-less spectral regions and was subtracted before the spectra were co-added. The reduced spectra are presented in Appendix\,\ref{appendix:spectra}. Finally, the spectral lines were fitted with Gaussians. The spectroscopic parameters of the observed transitions and results of the Gaussian fits are presented in Table\;\ref{lineparametres}. The error given in the integrated intensity includes the calibration uncertainty, which was taken to be $15\%$ at 2\,mm and $20\%$ at 1\,mm.

\begin{table*}
\caption{\label{lineparametres}Spectroscopic parameters of the observed lines.}
    \begin{tabular}{llllllll}
    \hline\hline
        Transition\tablefootmark{a} & Frequency & $A_\mathrm{ul}$\tablefootmark{b} & $E_\mathrm{up}$\tablefootmark{c} & $\int T_\mathrm{mb} dv$\tablefootmark{d} & $v_\mathrm{LSR}$\tablefootmark{e}  & $\Delta v$\tablefootmark{f}  & $T_\mathrm{peak}$\tablefootmark{g} \\
        $J_{K_a K_c}$ & MHz & s$^{-1}$ & K & K km s$^{-1}$ & km s$^{-1}$ & km s$^{-1}$ & K \\
\hline
\multicolumn{8}{c}{ortho$-$H$_2^{13}$CO}\\
\hline
$2_{1 2}-1_{1 1}$ & 137449.9503 & $4.93\times 10^{-5}$ & 21.7 & 0.331 (1) & 3.64 (0) & 0.48 (0) & 0.65\\
$2_{1 1}-1_{1 0}$ & 146635.6717 & $5.99\times 10^{-5}$ & 22.4 & 0.222 (1) & 3.61 (0) & 0.47 (0) & 0.44\\
$3_{1 3}-2_{1 2}$ & 206131.6260 & $2.11\times 10^{-4}$ & 31.6 & 0.166 (11) & 3.46 (1) &0.41 (3) & 0.38\\
$3_{1 2}-2_{1 1}$ & 219908.5250 & $2.56\times 10^{-4}$ & 32.9 & 0.095 (16) & 3.61 (5) & 0.62 (13) & 0.14\\
\hline
\multicolumn{8}{c}{para$-$H$_2^{13}$CO}\\
\hline
$2_{0 2}-1_{0 1}$ & 141983.7404 & $7.25\times 10^{-5}$ & 10.2 & 0.169 (2) & 3.44 (0) & 0.44 (1) & 0.36\\
$3_{0 3}-2_{0 2}$ & 212811.1840 & $2.61\times 10^{-4}$ & 20.4 & 0.041 (3) & 3.49 (1) & 0.27 (2) & 0.14 \\
\hline
\multicolumn{8}{c}{ortho$-$H$_2$C$^{18}$O}\\
\hline
$2_{1 2}-1_{1 1}$ & 134435.9203 & $4.61\times 10^{-5}$ & 21.5 & 0.039 (3) & 3.49 (2) & 0.39 (4) & 0.09\\
$2_{1 1}-1_{1 0}$ & 143213.0680 & $5.58\times 10^{-5}$ & 22.2 & 0.020 (0) & 3.57 (0) & 0.43 (1) & 0.04\\
\hline
\multicolumn{8}{c}{para$-$H$_2$C$^{18}$O}\\
\hline
$2_{0 2}-1_{0 1}$ & 138770.861  & $6.76\times 10^{-5}$ & 10.0 & 0.021 (2) & 3.63 (2) & 0.40 (5) & 0.05\\
\hline
    \end{tabular}
    \tablefoot{
    \tablefoottext{a} {Quantum numbers of the transition.}
    \tablefoottext{b} {Einstein $A$ coefficient.}
    \tablefoottext{c} {Energy of the upper level.}
    \tablefoottext{d} {Derived integrated intensity.}
    \tablefoottext{e} {Rest velocity.}
    \tablefoottext{f} {Full width at half maximum.}
    \tablefoottext{g} {Peak temperature.}
    }
\end{table*}

\section{Results}\label{section:results}
\subsection{Experiments}\label{section:experimentalresults}

The evolution of H$_2$CO ice infrared (IR) spectra as a function of fluence is depicted in Fig.~\ref{fig:IRspectra}. At fluence zero, the film is made of pure H$_2$CO, and neither paraformaldehyde nor CO or CO$_2$ can be seen in the spectra. As soon as irradiation starts, bands characteristic of CO$_2$ and CO appear, as well as bands around 2920, 1230, 1100, and 920 cm$^{-1}$, which we attribute to the formation of an H$_2$CO polymer (polyoxymethylene, hereafter POM). Conversely, H$_2$CO bands decrease with fluence. Apart from CO, CO$_2$, and POM, the infrared analysis shows no evidence of the formation of any other chemical species during the irradiation of H$_2$CO. The column densities of CO$_2$, CO, and H$_2$CO were derived using the band strengths given in Table\,\ref{IR bands assignments} and the integrated absorbance of the bands at 2343 cm$^{-1}$, 2139 cm$^{-1}$, and 1725 cm$^{-1}$, respectively. The POM column density was calculated using a combination mode at 1105 cm$^{-1}$ and a band strength of $1.9 \times 10^{-18}$\,cm\,molecules$^{-1}$ \citep{Noble12}. This band strength is not known with precision \citep{Tadokoro63,SCHUTTE1993118}, and we assumed a factor of 2 uncertainty. An additional uncertainty arises from changes in the refractive index value across strong absorbance bands, resulting in some unphysical negative absorbances after baseline correction (see e.g. the C=0 stretch of H$_2$CO). The derived column density evolution as a function of fluence, with uncertainty zones, is presented in Fig.~\ref{fig:CD_IR}. As can be seen, the POM column density is not accurately determined, but it is consistent with a fast and efficient polymerisation of \ce{H2CO} at very low fluence. Thus, at a fluence of $3 \times 10^{11}$ ions\,cm$^{-1}$, the ice is mainly composed of POM.

\begin{figure*}
\includegraphics[angle=0,width=\textwidth]{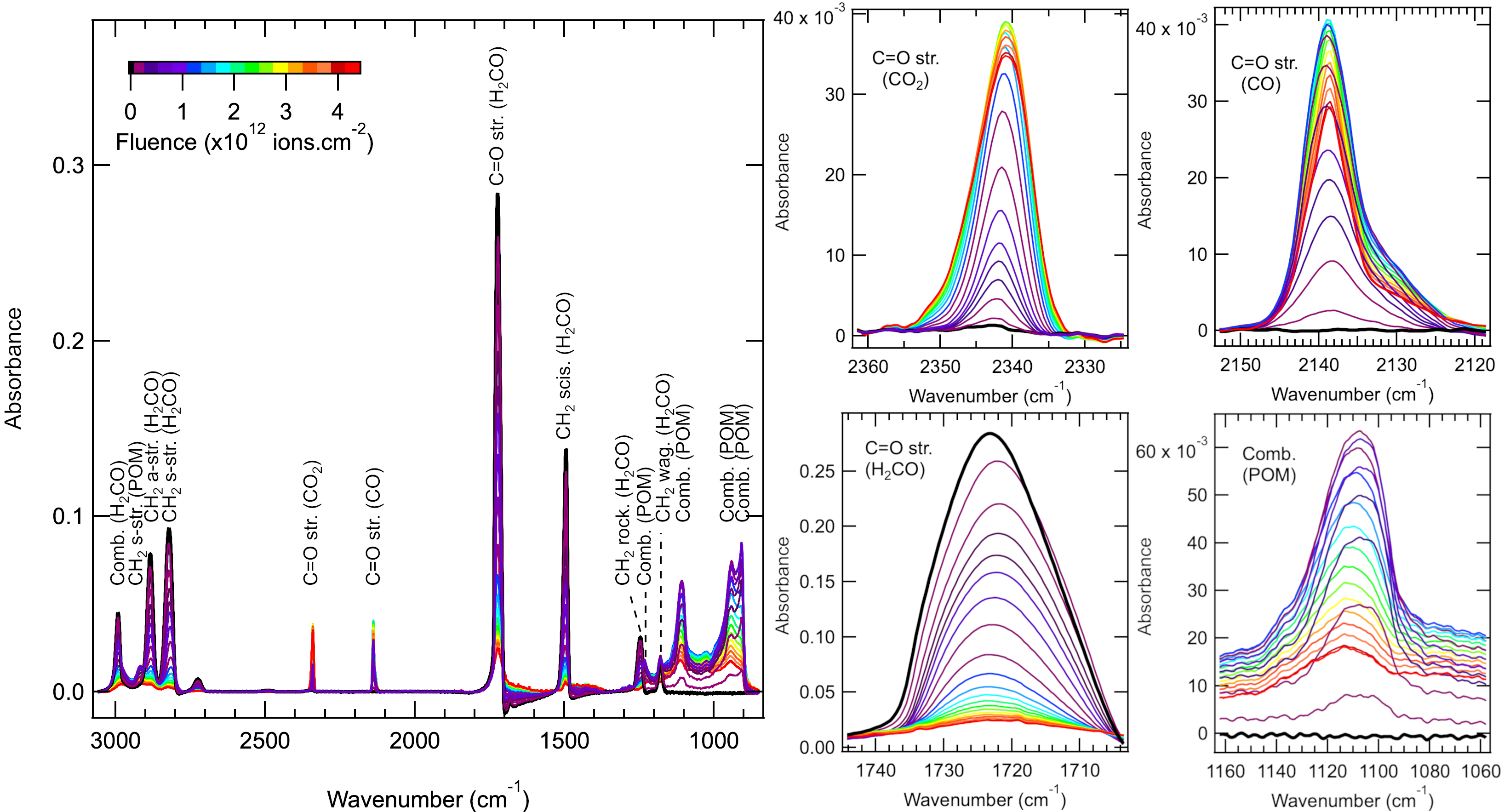}
\caption{IR spectra of irradiated H$_2$CO ice. \emph{Left panel}: Full range spectra, corrected for baseline, with assigned spectral peaks (see Table~\ref{IR bands assignments}). 
\emph{Right panel}: Focus on the evolution of the four bands used to derive column density of CO$_2$, CO, H$_2$CO, and POM in the ice.}
\label{fig:IRspectra}
\end{figure*}

\begin{table}
\caption{\label{IR bands assignments}
Peak position, assignments, and band strengths (at 25~K).}
\begin{tabular}{llcc}
\hline\hline
        Label & mode  & Position & $\mathcal{A}$  \\
              &       & cm$^{-1}$ & cm molecules$^{-1}$ \\
        \hline
        \multicolumn{4}{l}{H$_2$CO \tablefootmark{a}}\\
        \hline
        $\nu {_2} + \nu {_6}$  & comb. & 2997 & 3.2 $\times$ 10$^{-18}$\\
        $\nu {_5}$ & CH$_{2}$ a-str. & 2891 & 4.7 $\times$ 10$^{-18}$ \\
        $\nu {_1}$ & CH$_{2}$ s-str. & 2829 & 1.3 $\times$ 10$^{-17}$\\
        $\nu {_2}$ & C=O str. & 1725 & 1.6 $\times$ 10$^{-17}$\\
        $\nu {_3}$ & CH$_{2}$ scis. & 1500 & 5.1 $\times$ 10$^{-18}$ \\
        $\nu {_6}$ & CH$_{2}$ rock. & 1247 & 1.5 $\times$ 10$^{-18}$ \\
        $\nu {_4}$ & CH$_{2}$ wag. & 1178 & 7.2 $\times$ 10$^{-19}$ \\
        \hline
        \multicolumn{4}{l}{CO$_2$ \tablefootmark{a}}\\
        \hline
        $\nu {_3}$ & a-str. & 2343 & 7.6 $\times$ 10$^{-17}$ \\
        \hline
        \multicolumn{4}{l}{CO \tablefootmark{a}}\\
        \hline
        $1-0$ & str. & 2139 & 1.12 $\times$ 10$^{-17}$ \\
        \hline
        \multicolumn{4}{l}{POM \tablefootmark{b}} \\
        \hline
          & CH$_{2}$ s-str. & 2924 & - \\
          & comb. & 1235 & - \\
          & comb. & 1105 &  1.9 $\times$ 10$^{-18}$\\
          & comb. & 907-942 & - \\
        \hline
    \end{tabular}
    \tablefoot{
    \tablefoottext{a}{ \citet{Bouilloud15}.}
    \tablefoottext{b}{\citet{SCHUTTE1993118}.}
    }
\end{table}

\begin{figure}
\includegraphics[angle=0,width=\columnwidth]{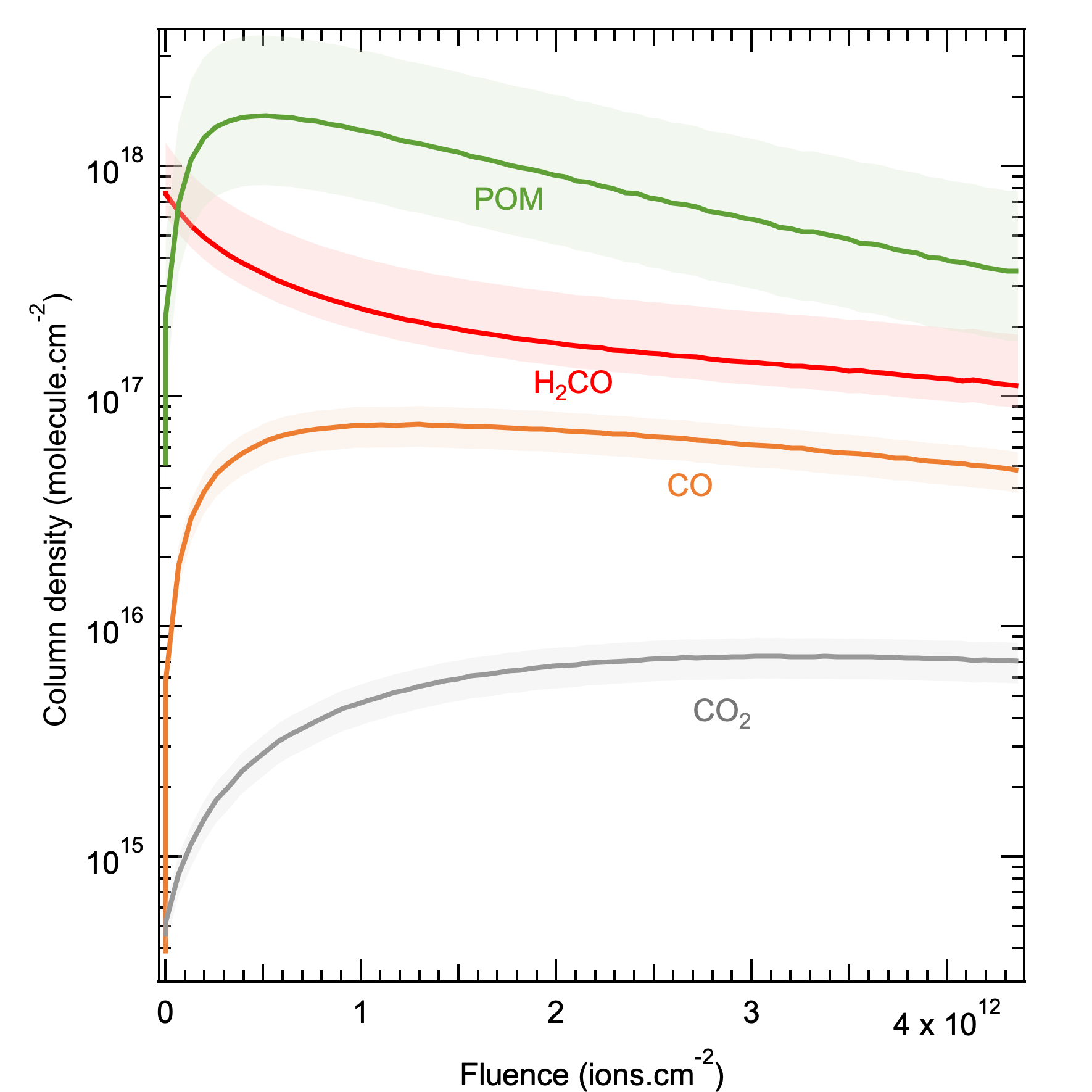}
\caption{Column density evolution of CO$_2$, CO, H$_2$CO, and POM ices as a function of fluence. In the case of POM, a factor of 2 uncertainty was assumed for the band strength of the combination mode at 1105 cm$^{-1}$  (see text), as depicted by the shaded green zone. In the case of H$_2$CO, the band strength of the stretching mode at 1725 cm$^{-1}$ was also evaluated by \cite{SCHUTTE1993118} as 9.6$\times$10$^{-18}$ cm\,molecules$^{-1}$ at 10~K, which corresponds to the highest column density value of the shaded red zone. Otherwise, an uncertainty of 20\% was assumed for the band strengths.}
\label{fig:CD_IR}
\end{figure}

In similar experiments on non-polymerisable ices such as \ce{CH3OH} and \ce{CO2} \citep{Dartois19,seperuelo09}, the column density evolution with fluence can be solved through a simple differential equation involving solely radiolysis and sputtering processes \citep[see e.g. Eq.~(3) in ][]{Dartois19}. Here, however, we were not able to solve such a differential equation, due to the additional polymerisation process of \ce{H2CO} towards POM, whose exact nature is unknown.\footnote{In particular, the number of monomers is unknown and most probably the synthesised POM is a mixture of polymers of various chain sizes.} In addition, while swift ion irradiation of \hhco is responsible for the formation of POM, it also destroys POM to CO and CO$_2$ (see the decrease in POM column density beyond a fluence of $\sim 3\times 10^{11}$ ions cm$^{-2}$ in Fig.\,\ref{fig:CD_IR}). Therefore, in order to derive the sputtering yield, we resorted to another method based on the determination of the reduction in thin film thickness using the Fabry–Perot interference fringes observed in the infrared spectra \citep{Dartois20N2, Dartois23}, combined with QMS analysis of the gas composition. We emphasise that in order to validate this method, it was first applied to measure the thickness evolution of an irradiated thin film of CO$_{2}$, for which the usual fit of the column density evolution via a differential equation was also possible. As reported in Appendix~\ref{appendix:co2}, both methods were found to give consistent results, which are also in good agreement with the literature.

In the case of \ce{H2CO}, as irradiation proceeds, the ice film becomes a mixture of mainly POM and \hhco for which the refractive index (\(n\)) is not available. Fitting the rigorous expression for the transmission (\(T\)) of a thin absorbing film on a thick transparent substrate is therefore not possible. Nevertheless, transmittance spectra follow a sinusoidal pattern, where the period is proportional to the thickness. Thus, we used the following simple expression to fit the transmittance spectra in the ranges of wavenumbers ($1/\lambda$) free of molecular absorptions:
\begin{equation}
\label{eq:interference}
    T = T_0 + A \sin{\left(\frac{4\pi n d}{\lambda \cos(12 ^\circ)}\right)}.
\end{equation}
Here $T_O$ and $A$ are free parameters, $n$ is the refractive index taken as $n=1.33\pm0.04$ \citep{Bouilloud15}, and \(d\) is ice thickness. The factor $\cos(12^\circ)$ corrects for the angle between the infrared beam and the sample. Fits obtained with this expression are shown in Fig.~\ref{fig:Transmittance}. The ice film is composed of POM and H$_2$CO, with a composition changing during irradiation. Therefore, evaluating the sample mass density $\rho$ is challenging. Furthermore, even the mass density of a pure \hhco\ ice film is unknown. Consequently, it is assumed to be constant at the value for liquid formaldehyde, $\rho = 0.81\,\text{g}\,\text{cm}^{-3}$ \citep{Bouilloud15}, with an assumed uncertainty of 20\%. It was used to derive the surface density of the ice \(\sigma = \rho d\), as a function of fluence, as shown in Fig.\,\ref{fig:MassPerArea}. The total mass sputtering yield can be directly estimated from the slope of the surface density evolution with fluence as \(Y_{m, tot} = (1.8\pm0.5) \times 10^{-18}\) g\,ion$^{-1}$. It should be noted that this total yield includes all sputtered species from the considered volume,  both radiolysed species (C, O, H, H$_2$, CO, and CO$_2$) and intact \ce{H2CO} fragments.


\begin{figure}
\includegraphics[angle=0,width=\columnwidth]{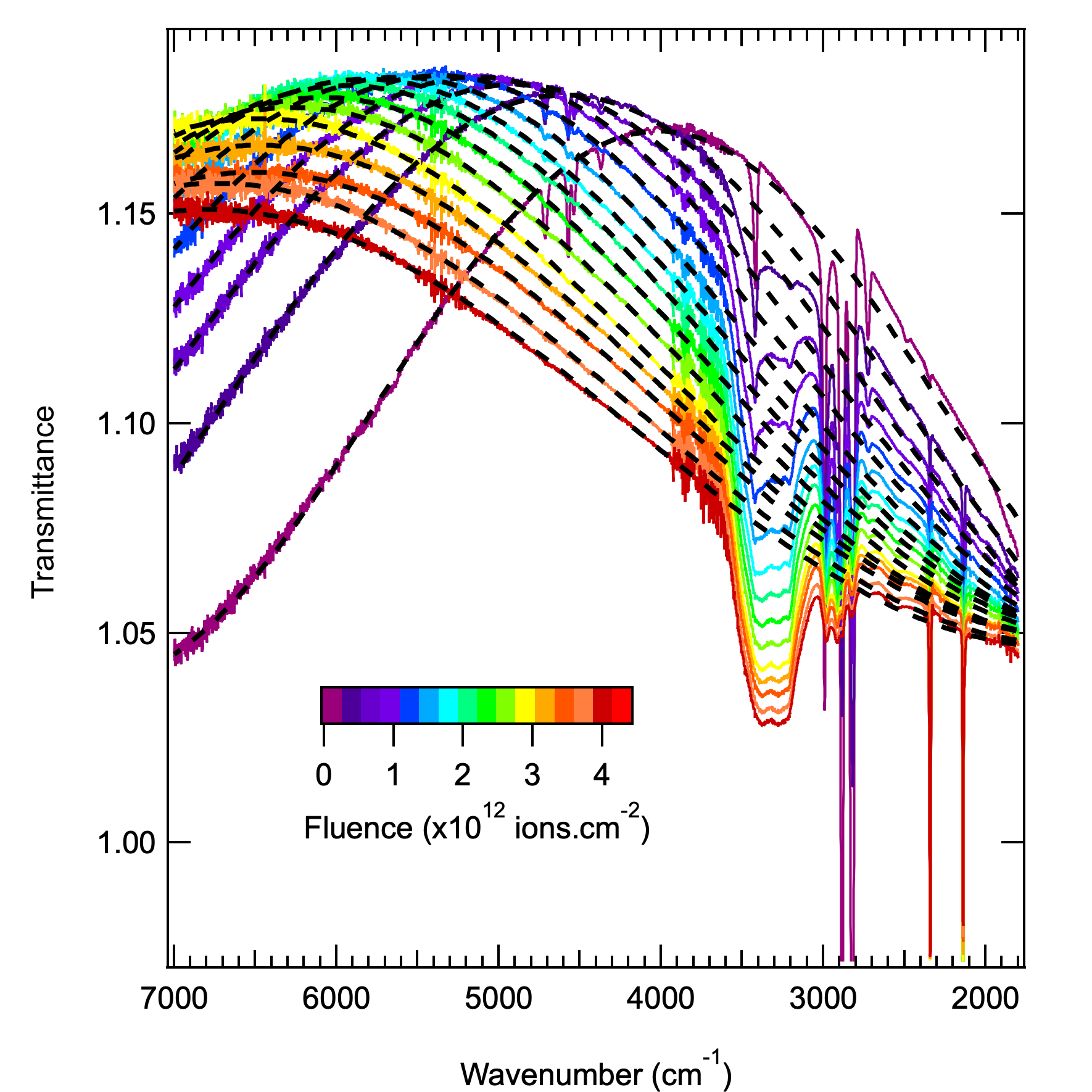}
\caption{
 Infrared transmittance spectra of a H$_{2}$CO ice film evolution as a
function of fluence. \emph{Solid lines}: Experimental data. \emph{Dashed lines}: Model spectra fitted to the
data. See text for details.}
\label{fig:Transmittance}
\end{figure}

\begin{figure}
\includegraphics[angle=0,width=\columnwidth]{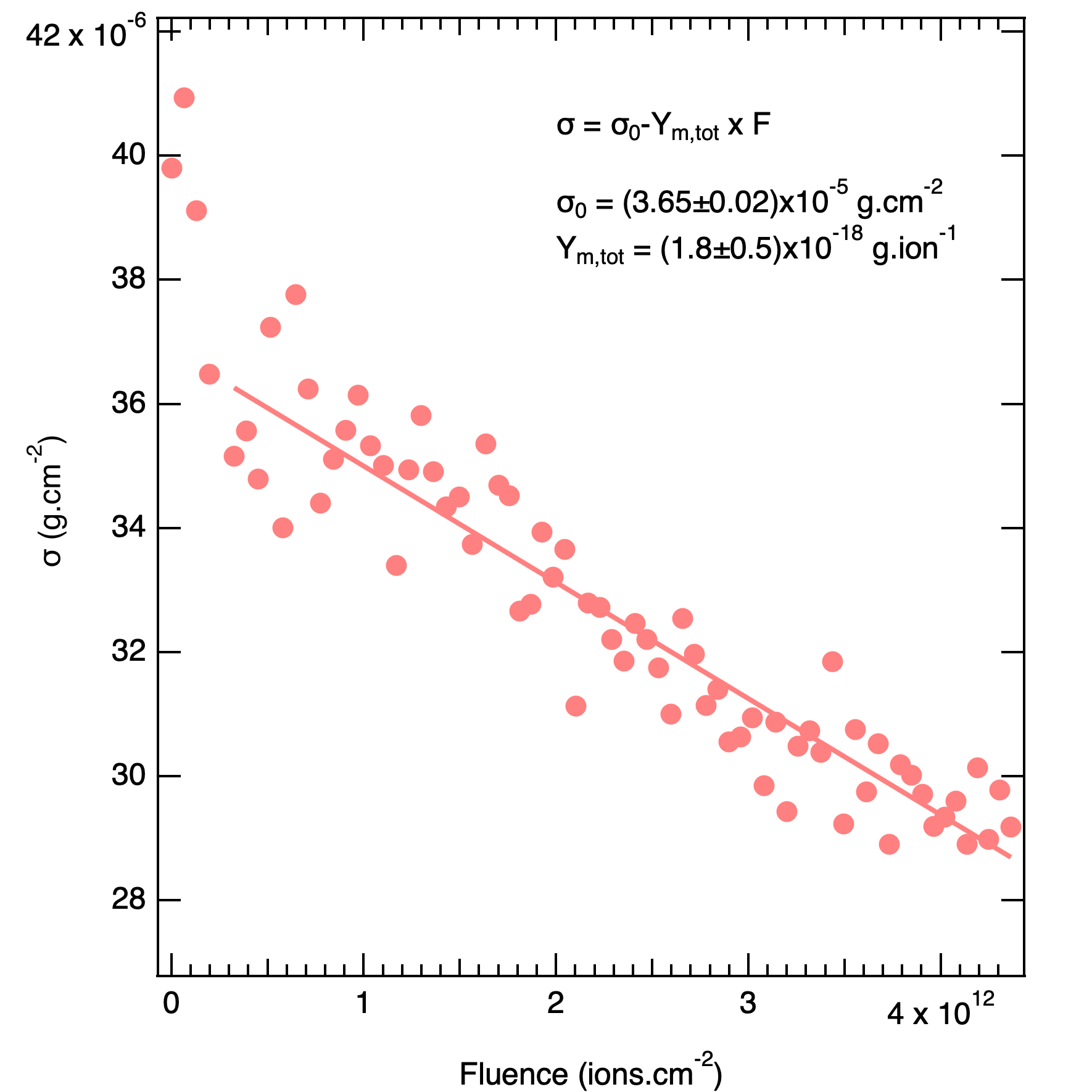}
\caption{Evolution of the surface density as a function of fluence deduced from interference fringes. The fit was performed for fluences greater than 6$\times 10^{11}$ ions\,cm$^{-2}$ since the very beginning of the irradiation corresponds to ice compaction \citep[e.g.][]{Palumbo2006, Dartois2013}.}
\label{fig:MassPerArea}
\end{figure}

In order to determine the chemical composition of the sputtered mass volume, the various fragments ejected into the gas phase were measured with the QMS. The small residual or background signal observed before irradiation was subtracted from the recorded signal during the whole irradiation time, as reported in Fig.~\ref{fig:QMS} (top panel). We note that there is no intact sputtered POM as no signal is observed at $m/z$=31, a mass characteristic of POM fragmentation pattern \citep{butscher19, duvernay14}. The mass $m/z$=30 can be used to follow \hhco molecules since no fragment at this mass is observed in the CO fragmentation pattern (Fig.~\ref{fig:Fragmentation_Pattern}). Thus, the signals at m/z=30 and m/z=28 can be used to retrieve the total number of \hhco and CO species, respectively, by dividing each m/z signal by its relative contribution in the fragmentation pattern. The remaining signals at m/z=12 and m/z=16, after subtracting the contributions of CO and H$_2$CO, indicate the abundances of C and O atoms, respectively. Once corrected for the total electron-impact ionisation cross-section \(\sigma^{impact}(X)\) at 70~eV (energy of the QMS electron ionisation source), the relative fraction of each species \(f(X,g)\)  with fluence $F$ is determined (Fig.~\ref{fig:QMS}, bottom panel). QMS measurements at very low masses are not reliable, but the amount of hydrogen (H or H$_2$) is supposed to be equal to the sum of the amounts of C, O, and CO. We note that CO$_2$, a minor component of the ice (see Fig. \ref{fig:IRspectra}), was not detected in the gas phase by the QMS. As irradiation proceeds, the composition of the film changes (Fig. \ref{fig:CD_IR}) and this is also observed in the gas composition (Fig. \ref{fig:QMS}, bottom panel). The fraction of gaseous H$_2$CO thus decreases significantly (by approximately 70 \%) for initial fluences (up to 2$\times 10^{11}$ ions\,cm$^{-2}$), which is consistent with the radiolysis and the rapid polymerisation of solid \ce{H2CO}. 
It then decreases more gradually, before a (noisy) slightly ascending plateau is reached beyond fluences of 3$\times 10^{12}$ ions\,cm$^{-2}$. Conversely, the fractions of C and O increase rapidly (by approximately 70\%), but also reach a plateau. Overall, the molecular fractions of these three chemical species in the gas vary by less than a factor of two. The larger molecular fractions of CO and H$_2$ increase up to fluences of about 2$\times 10^{12}$ ions\,cm$^{-2}$ before reaching a slightly descending plateau, with variations of less than 15\%. We can extract the initial fraction of each species $X$ in the gas phase \(f_{i}(X,g)\) (Table~\ref{tab:yields}), and the subsequent evolution \(f(X,g)\) was fitted by third-order polynomials.

\begin{figure}
\includegraphics[angle=0,width=\columnwidth]{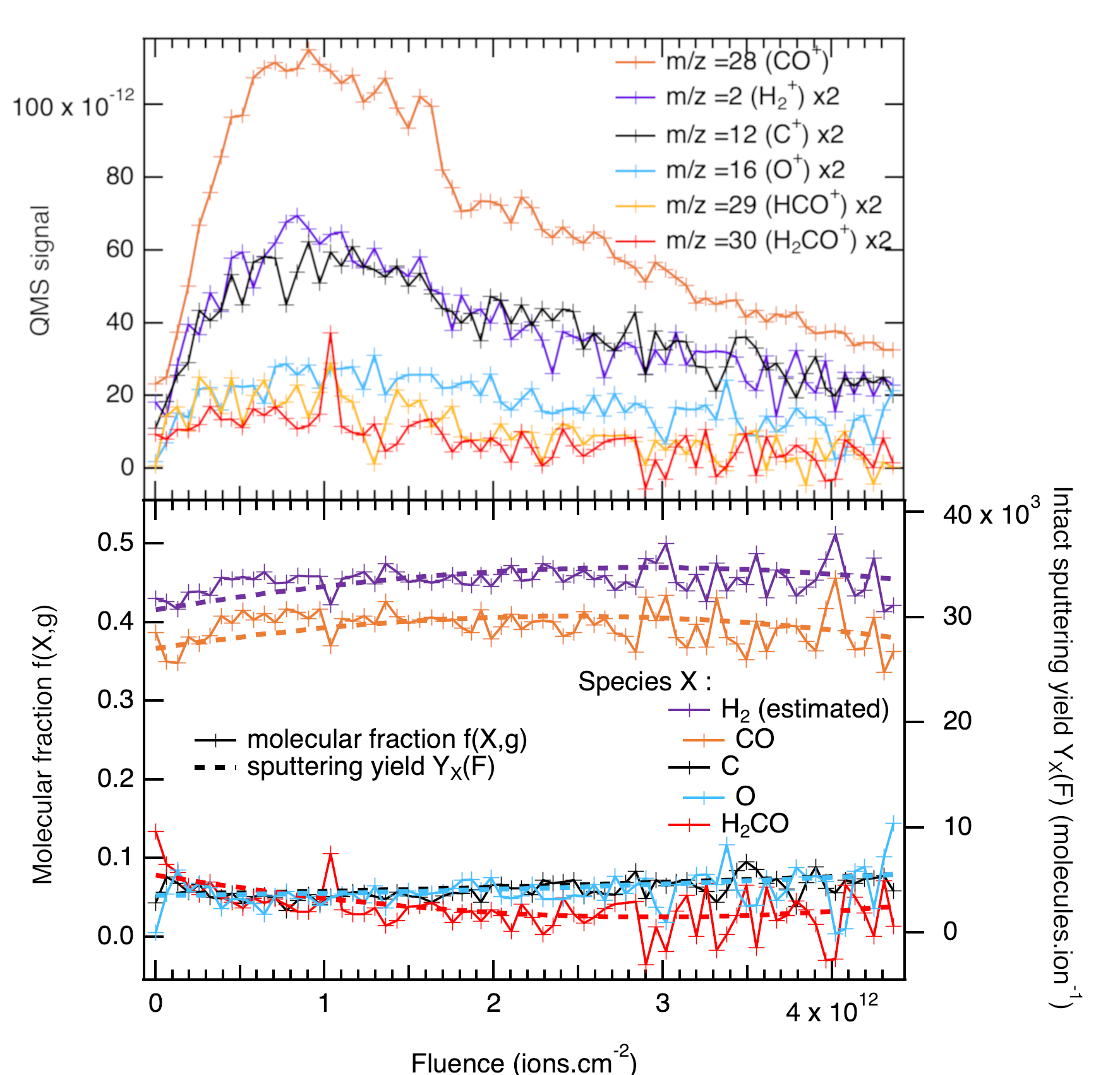}
\caption{
	QMS signals as a function of fluence. \emph{Top panel}: Background corrected signals. \emph{Bottom panel}: Molecular fractions in the gas (\emph{solid lines}, left $y$-axis) derived from the fragmentation patterns and ionisation cross sections, and sputtering yield of each species (\emph{dashed lines}, right $y$-axis). See text for details.}
\label{fig:QMS}
\end{figure}

\begin{figure}
\includegraphics[angle=0,width=\columnwidth]{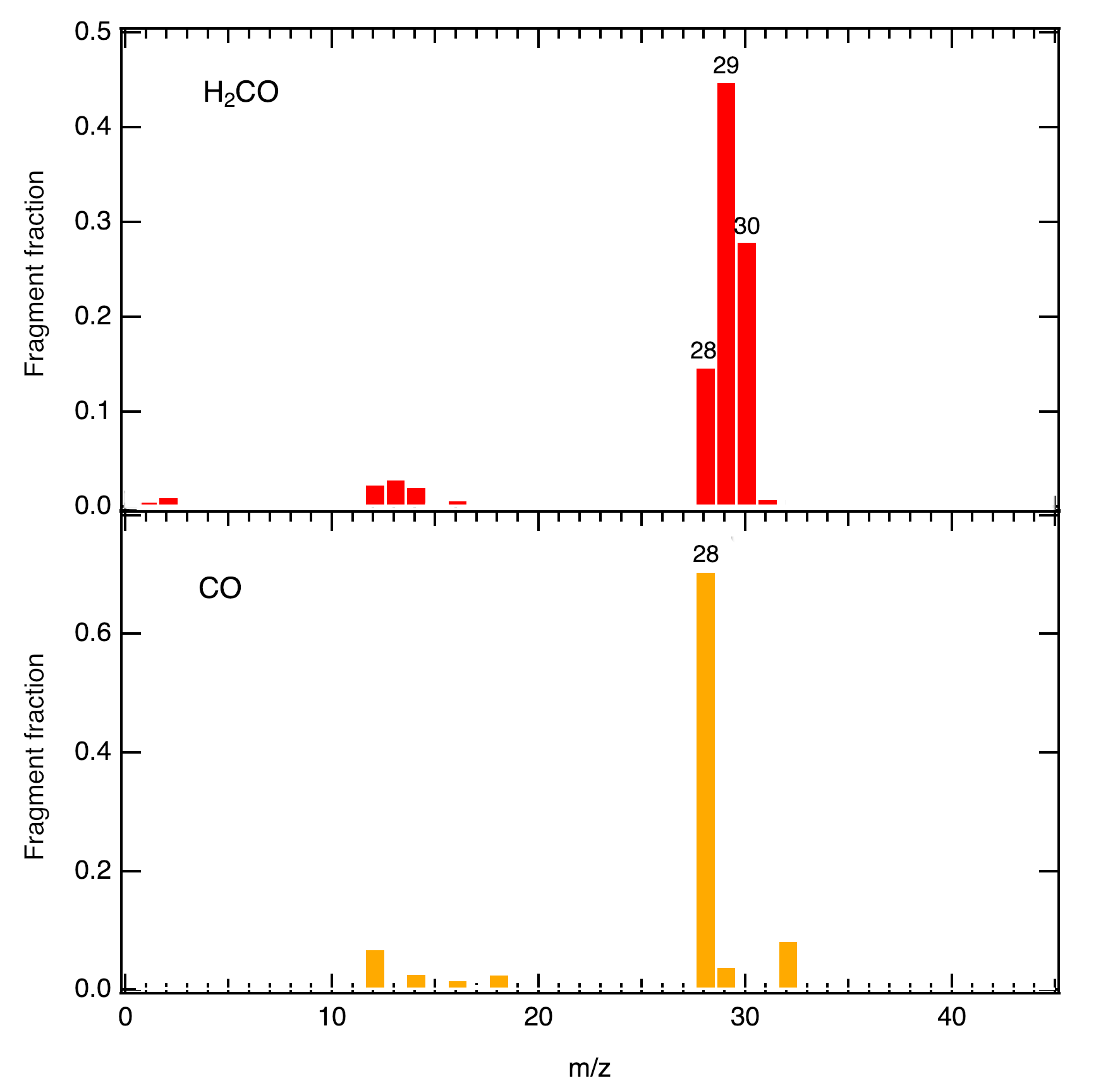}
\caption{Mass fragmentation pattern monitored during species injection (for ice sample deposition) for H$_{2}$CO (\emph{top panel}) and CO (unpublished data) (\emph{bottom panel}). }
\label{fig:Fragmentation_Pattern}
\end{figure}

By combining the total mass sputtering yield \(Y_{m, tot}\) with the fitted molecular fractions \(f(X,g)\), we can derive the sputtering yield of each ejected species \(Y_{X}(F)\),

\begin{equation}
\label{eq:Yxintact}
    Y_{X}(F) = Y_{m,tot} \times f_m(X,g) \times \frac{N_A}{M(X)},
\end{equation}
where \(N_A\) is the Avogadro constant, \(M(X)\) is the molecular mass of species \(X\), and \(f_{m}(X,g)\) represents the mass fraction of species \(X\) in the gas:
\begin{equation}
\label{eq:fm}
    f_{m}(X,g) = \frac{f(X,g) \times M(X)}{\sum_{X}(f(X,g) \times M(X))}.
\end{equation}

The derived sputtering yields for the various ejected species $X$ are plotted as dashed lines in the bottom panel of Fig.~\ref{fig:QMS} (right $y$-axis). Similarly to the molecular fractions, the intact sputtering yields vary by less than a factor of 2 for C, O, and H$_{2}$CO, and by less than 15\% for CO and H$_{2}$, as expected. The average intact sputtering yields $Y_{X}$ were computed over the full range of fluences and are listed in Table~\ref{tab:yields}. The average sputtering yield for intact \ce{H2CO} is found to be $Y_{intact}=Y_{\rm H_2CO}=(2.5\pm 2.3)\times 10^3$ molecules\,ion$^{-1}$.

\begin{table}
\caption{\label{Species sputtering yield} Total electron-impact ionisation cross-sections $\sigma^{impact}(X)$; initial fractions $f_{i}(X,g)$ and averaged sputtering yield $Y_{X}$.}
\begin{tabular}{cccc}
\hline\hline
        Species (X) & $\sigma_{impact}$\tablefootmark{a} & $f_{i}(X,g)$ &$Y_{X}$ \\
        & $\AA^2$ & & molecules\,ion$^{-1}$\\
        \hline
        H$_{2}$&1.021& 0.43 & (3.4$\pm$1.0)$\times$10$^4$ \\
        CO &2.516&0.39 & (2.9$\pm$0.8)$\times$10$^4$ \\
        C&2.317& 0.04 & (4.5$\pm$4.1)$\times$10$^3$\\
        O&1.363& 0.01 & (4.3$\pm$3.9)$\times$10$^3$\\
        H$_{2}$CO &4.140& 0.13 & (2.5$\pm$2.3)$\times$10$^3$ \\
        \hline
    \end{tabular}
    \tablefoot{
    \tablefoottext{a}{From the NIST database (https://webbook.nist.gov/chemistry/), except $\sigma^{impact}(\rm{C})$ and $\sigma^{impact}(\rm{O})$ from \cite{Kim02}.}
    }
\end{table}

We can now try to estimate the effective sputtering yield, $Y_{eff}$. For non-polymerisable species, $Y_{eff}$ is usually determined using the column density evolution with fluence measured from IR spectroscopy \citep{Dartois19,seperuelo09}. For a pure ice $X$, $Y_{eff}$ is defined as the semi-infinite thickness effective-sputtering  yield for the ice under study. It includes both intact sputtered species $X$ and radiolysed (daughter) sputtered species. Since radiolysis and sputtering occur simultaneously, IR spectroscopy captures the disappearance of H$_2$CO, regardless of whether it is sputtered intact or radiolysed. $Y_{eff}$ is simply related to $Y_{intact}$ as
\begin{equation}
\label{eq:Yeff}
Y_{eff}=\frac{Y_{intact}}{\eta},
\end{equation}
where $\eta$ is the intact fraction of sputtered H$_2$CO.
In our experiment, the volume to be sputtered just before irradiation is composed of pristine \ce{H2CO} ice. The impact of a swift heavy ion induces radiolysis along its track, affecting a fraction of this volume. Therefore, after irradiation has started, the sputtered volume is composed of intact \ce{H2CO}, fresh products of radiolysis as well as primary radiolytic products that accumulate in the film (see Fig.~\ref{fig:Dessin_Film_evolution}). Therefore, in order to determine the effective sputtering yield $Y_{eff}$, the intact fraction ($\eta$) of sputtered \hhco during irradiation must be estimated using the initial fractions reported in Table~\ref{Species sputtering yield}:
\begin{equation}
\label{eq:eta}
    \eta = \frac{f_i({\rm H_2CO},g)}{f_i({\rm H_2CO},g)+ f_i({\rm CO},g)+f_i({\rm C},g)}
.\end{equation}

\noindent In practice, we derived an intact fraction of sputtered \ce{H2CO} molecules $\eta$=0.23$\pm 0.03$, corresponding to an effective sputtering yield for the \ce{H2CO} film of $Y_{eff}$=(10.9$\pm$9.9)$\times$10$^{3}$ molecules\,ions$^{-1}$. 

\begin{figure*}
\includegraphics[angle=0,width=\textwidth]{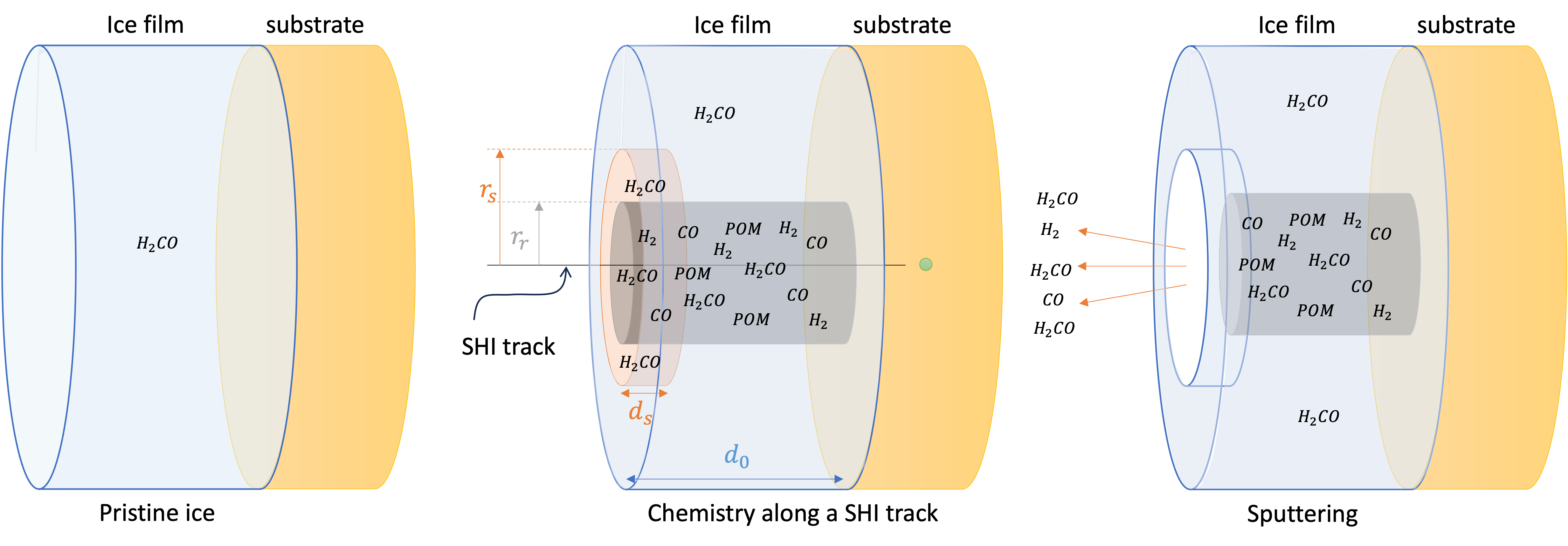}
\caption{Schematic timeline of events during the first impact of a swift heavy ion (SHI) on an ice sample, within the  thermal spike  model. Chemistry and sputtering occur almost simultaneously within 10$^{-9}$~seconds of the ion's passage \citep{Lang20}. The length $d_{s}$ corresponds to the sputtering depth probed by an individual ion incident, which is inferior to $d_{0}$, the initial ice film thickness; $r_{r}$ corresponds to the radiolysis destruction radius, which is at most equal to the sputtering cylinder radius $r_{s}$ \citep{Dartois21}. Considering the timescales involved, infrared spectroscopy cannot observe the radiolysis effects within the sputtered volume during the passage of an ion. 
 }
\label{fig:Dessin_Film_evolution}
\end{figure*}


In their compendium of selected literature sputtering yields, \cite{Dartois23}  show  that 
the sputtering yield induced by swift heavy ions follows an empirical law that is directly proportional to the square of the electronic stopping power $S_e$:
\begin{equation}\label{eq:Ye}
    Y_{intact}=Y_0S_e^2
.\end{equation}
Here, using a stopping power of 2830~eV\,(10$^{15}$\,molecules\,cm$^{-2}$)$^{-1}$ for the $^{86}$Kr$^{18+}$ beam, we obtain for a pure \ce{H2CO} ice  $Y_0$=(3.1$\pm$2.8)$\times$10$^{-4}$\,10$^{30}$\,eV$^{-2}$\,cm$^{-4}$\,molecule$^{3}$. 
However, most studies only have access to the effective sputtering yield $Y_{eff}$, which is therefore used for $Y_{intact}$ in Eq. \ref{eq:Ye} \citep[e.g.][]{Dartois23}. In this case, the effective prefactor value is $Y_0^{eff}$=(1.4$\pm$ 1.3)$\times$10$^{-3}$\,10$^{30}$\,eV$^{-2}$\,cm$^{-4}$\,molecule$^{3}$.

\subsection{H$_2$CO abundance in L1689B}

The H$_2$CO column density in the L1689B prestellar core was determined using the RADEX radiative transfer code \citep{2007A&A...468..627V} in the uniform sphere approximation. The collisional rate coefficients used in the computation are the H$_2$CO$-$H$_2$ coefficients from \citet{2013MNRAS.432.2573W} adapted to the spectroscopy of H$_2$C$^{18}$O and H$_2^{13}$CO and were obtained from the Excitation of Molecules and Atoms for Astrophysics (EMAA) collisional rate database.\footnote{emaa.osug.fr} Because there are no radiative or collisional mechanisms converting one form into the other, the ortho and para nuclear-spin isomers of each isotopologue were treated as two separate species. At thermal equilibrium, the ortho-to-para ratio of \hhco is $\sim $ 1.5 at 10~K, and reaches the statistical value of 3.0 at about 30~K. 
Only collisions with para-H$_2$ were considered as the ortho-H$_2$ abundance represents at most $1-10\%$ of the total H$_2$ \citep{2009A&A...494..623P,Troscompt09,2012A&A...537A..20D} and can be neglected in prestellar cores.

Using RADEX, we calculated grids of models for ranges of values in kinetic temperature, H$_2$ density, and in the molecular column densities (ortho-H$_2^{13}$CO, para-H$_2^{13}$CO, ortho-H$_2$C$^{18}$O, or para-H$_2$C$^{18}$O). The grid points in temperature and H$_2$ density were chosen to be the same for all species, which is based on the assumption that all of them are coexistent. The densities span a range between $5\times 10^4$\,cm$^{-2}$ and $5\times 10^5$\,cm$^{-2}$ and the temperatures between 3\,K and 12\,K. The column density of ortho-H$_2^{13}$CO $N_\mathrm{o13}$ was varied between $5\times 10^{11}$\,cm$^{-2}$ and  $5\times10^{12}$\,cm$^{-2}$. The column density of para-H$_2^{13}$CO $N_\mathrm{p13}$ was taken as $N_\mathrm{p13}=N_\mathrm{o13}/opr$ where $opr$ is the ortho-to-para ratio in the considered H$_2$CO isotopologue (either in H$_2^{13}$CO or in H$_2$C$^{18}$O), and the free parameter $opr$ was varied between 1 and 6. For the column density of H$_2$C$^{18}$O, we assumed a ratio $cor= \,^{13}\mathrm{C}/^{18}\mathrm{O}$  such as the ortho-H$_2$C$^{18}$O column density $N_\mathrm{o18}$ is $N_\mathrm{o18}=N_\mathrm{o13}/cor$ and the para-H$_2$C$^{18}$O column density $N_\mathrm{o18}$ is $N_\mathrm{p18}=N_\mathrm{o13}/(cor\times opr)$, and varied this ratio between 7 and 10 \citep[see e.g.][for standard values of $^{13}$C/$^{18}$O]{1999RPPh...62..143W}. The input linewidths for the calculations were taken to be 0.48\,km\,s$^{-1}$ for H$_2^{13}$CO and 0.38\,km\,s$^{-1}$ for H$_2$C$^{18}$O, but they have little influence on the results, because our fitting procedure relies on the velocity integrated intensities of the modelled lines compared to that of the observations. 
The goodness of fit was estimated with a $\chi^2$ defined as
\begin{equation}
\begin{split}
\chi^2(N_\mathrm{o13},T,n_\mathrm{H_2},opr,cor) = \sum_{i}\frac{(W_i^\mathrm{mod}(N_\mathrm{o13},T,n_\mathrm{H_2})-W_i^\mathrm{obs})^2}{\sigma_i^2}\\
+\sum_{j}\frac{(W_j^\mathrm{mod}(N_\mathrm{o13},T,n_\mathrm{H_2},opr)-W_j^\mathrm{obs})^2}{\sigma_j^2}\\
+\sum_{k}\frac{(W_k^\mathrm{mod}(N_\mathrm{o13},T,n_\mathrm{H_2},cor)-W_k^\mathrm{obs})^2}{\sigma_k^2}\\
+\sum_{l}\frac{(W_l^\mathrm{mod}(N_\mathrm{o13},T,n_\mathrm{H_2},opr,cor)-W_l^\mathrm{obs})^2}{\sigma_l^2}
\end{split}
,\end{equation}
where $T$ is the kinetic temperature; $n$(H$_2$) is the H$_2$ density; $W^\mathrm{obs}$ is the observed line integrated intensity; $\sigma$ is the error on the observed integrated intensity; $W^\mathrm{mod}$ is the line integrated intensity calculated by RADEX; and the indices $i,j,k,l$ refer to the transitions in ortho-H$_2^{13}$CO, para-H$_2^{13}$CO, ortho-H$_2$C$^{18}$O, and para-H$_2$C$^{18}$O, respectively. 
Typically, 10 to 15 points were chosen for each parameter. The grids were refined close to parameter values best reproducing the observations.
The best fitting parameters are given in Table\,\ref{h2co_coldens}. 

\begin{table}
\caption{\label{h2co_coldens}
Non-LTE model parameters best fitting the observed spectra.}
\begin{tabular}{llllll}\hline\hline
        $T$\tablefootmark{a} & $n_{\rm H_2}$\tablefootmark{b} & $N_\mathrm{o13}$\tablefootmark{c} & $opr$\tablefootmark{d} & $cor$\tablefootmark{e} & $\chi^2$\\
        (K) & (cm$^{-3}$) & (cm$^{-2}$) & & & \\
        \hline
        10 & $2\times 10^5$ & $2.0\times 10^{12}$ & 3.6 & 9.1 & 4.3 \\
        \hline
    \end{tabular}
    \tablefoot{
    \tablefoottext{a}{Temperature.}
    \tablefoottext{b} {H$_2$ density.}
    \tablefoottext{c} {ortho-H$_2^{13}$CO column density.}
    \tablefoottext{d}{ortho-to-para ratio for H$_2$CO isotopologues.}
    \tablefoottext{e}{$^{13}$C/$^{18}$O ratio.}
    }
\end{table}

Because the ortho-to-para ratio is expected to be at most 3 for H$_2$CO,\footnote{A recent laboratory experiment has shown that the measured ortho-to-para ratio of \ce{H2CO} molecules formed via UV photolysis of \ce{CH3OH} ices is $\sim 3$, with no correlation with ice temperature \citep{Yocum23}.} we   re-ran the model and  kept the ortho-to-para ratio set to 3. In this case, the $\chi^2$ value is slightly worse (5.5 instead of 4.3 for the best model), and the best fitting temperature is slightly lower (closer to 9\,K). The other parameters (density, ortho-H$_2^{13}$CO column density, and $^{13}$C/$^{18}$O ratio) remain unchanged. We  also ran the fitting routine with the temperature and density values found for CH$_3$OH by \citet{2016A&A...587A.130B} (i.e. $T=8.5$\,K and $n_{H_2}=3.6\times 10^{5}$\,cm$^{-3}$), and again found  results for the parameters very similar to the best fit: $N_\mathrm{o13} = 1.8\times 10^{12}$\,cm$^{-2}$.  

The line integrated intensities from the best fit model are shown in Fig.\,\ref{fig:fit} against the observed line integrated intensities. The model can account for the observations within the uncertainties.

\begin{figure}
\includegraphics[angle=0,width=\columnwidth]{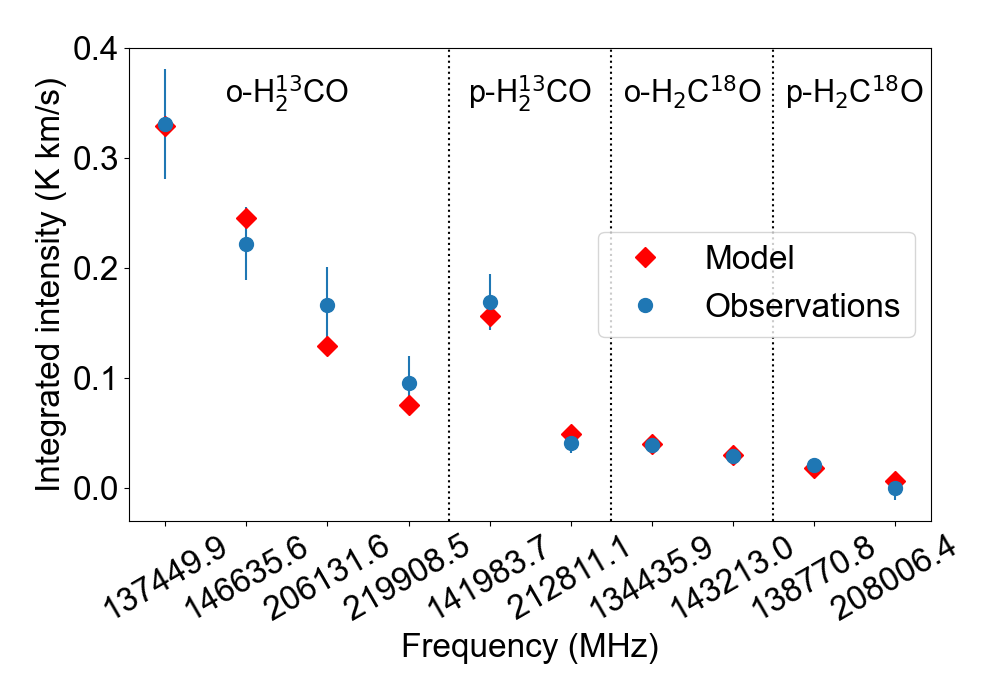}
\caption{Comparison between the observed integrated intensities (blue dots) and the integrated intensities derived from the non-LTE model (red diamonds), for each of the observed lines, labelled by their rest frequencies. In the figure, `o' stands for ortho and `p' stands for para.}
\label{fig:fit}
\end{figure}

With the values in Table\,\ref{h2co_coldens}, we obtain the following values for the molecular column densities: $N_\mathrm{o13}=2.04\times 10^{12}$\,cm$^{-2}$, $N_\mathrm{p13}=5.4\times 10^{11}$\,cm$^{-2}$, $N_\mathrm{o18}=2.2\times 10^{11}$\,cm$^{-2}$, and $N_\mathrm{p18}=6\times 10^{10}$\,cm$^{-2}$. Assuming a ratio $^{16}\mathrm{O}/^{18}\mathrm{O}$ of 557 \citep{1999RPPh...62..143W}, we obtain a total (ortho $+$ para) column density of $1.6\times 10^{14}$\,cm$^{-2}$ for the main isotopologue H$_2$CO.

The H$_2$ column density at the same position as the H$_2$CO observations was derived from the model of \citet{2016A&A...593A...6S}, by integrating the density profile along the line of sight and multiplying by two to account for the contribution of the far side of the core. The uncertainty was determined using the error bars from the fit including the interstellar radiation field \citep[see][for details]{2016A&A...593A...6S}. We find a total H$_2$ column density $N_{H_2} = 1.3\times 10^{23}$\,cm$^{-2}$, with a spread $5\times 10^{22} - 2.1\times 10^{23}$\,cm$^{-2}$. The corresponding relative abundance of \ce{H2CO} with respect to total hydrogen ($n_{\rm H}=n({\rm H}) + 2n({\rm H_2}))$ is $\sim 6\pm 4\times 10^{-10}$. 

\section{Astrophysical applications}\label{section:modelling}

\subsection{Model}

We investigate in this section whether non-thermal desorption from icy grains can explain the gas-phase abundance of \ce{H2CO} derived in L1689B. A simple analytical model is employed: \hhco molecules from the gas phase can only accrete on the icy surface of dust grains, while \hhco molecules from the grain surface can only desorb via photodesorption (due to both external and internal cosmic-ray induced UV photons) and direct cosmic-ray induced sputtering. 

In addition, a steady-state is assumed, with constant kinetic temperature, density, and total \hhco abundance (gas and ice) throughout the core. This total \hhco abundance is unknown in L1689B and taken as $3\times 10^{-6}$, which is the typical \hhco ice abundance in low-mass young stellar objects (LYSOS) \citep{2015ARA&A..53..541B}, corresponding to about 6\% with respect to H$_2$O ice.\footnote{This value is similar to the 4.1\% recently measured by JWST towards the young-mass protostar NGC~1333~IRAS 2 \citep{2024A&A...681A...9R}.} The kinetic temperature $T$ is fixed at 10~K and the H$_2$ density at $n(\rm H_2)=2\times 10^5$~cm$^{-3}$, as reported in Table\,\ref{h2co_coldens} for the L1689B core.           

At steady-state, the rates of accretion and desorption are equal and we can write:
\begin{equation}
    R_{acc}n^g_{f}= R_{des}n^s_{f},
    \label{ss}
\end{equation}
where $n^g_{f}$ and $n^s_{f}$ are the formaldehyde densities (in cm$^{-3}$) in the gas and solid ice phase, respectively. The accretion rate (in s$^{-1}$) can be written as
\begin{equation}
    R_{acc} = \left<\pi a_g^2\right>n_gv_{th}S,
    \label{racc}
\end{equation}
where $\left<\pi a_g^2\right>$ is the average grain surface (in cm$^{2}$), $n_g$ the grain number density (cm$^{-3}$), $v_{th}$ the thermal velocity of \hhco molecules, and $S$ the sticking coefficient taken as 1.0. The desorption rate (in s$^{-1}$) can be written as \citep{Flower2005}
\begin{equation}
    R_{des} = \frac{\Gamma^{tot}\left<\pi a_g^2\right>n_g}{\sum_i n^s_i},
\label{rdes}
\end{equation}
where $\sum_i n^s_i$ is the sum of ice abundances such that $n^s_f/\sum_i n^s_i$ is the fractional abundance of \ce{H2CO} in the ice. In the following, we   note that $\sum_i n^s_i=X_{\rm ices}n_{\rm H}$. The quantity 
$\Gamma^{tot}$ is the total desorption rate (in molecules~cm$^{-2}$s$^{-1}$) defined as
\begin{equation}
    \Gamma^{tot} = \Gamma_{ISRF} + \Gamma_{CRI-UV} + \Gamma_{CRD}, 
\end{equation}
where $\Gamma_{ISRF}$, $\Gamma_{CRI-UV}$, and $\Gamma_{CRD}$ are the desorption rates due to external UV photodesorption, cosmic-ray induced UV phodesorption, and direct cosmic-ray sputtering, respectively. We note that cosmic-ray induced whole-grain heating can be neglected for a thick volatile mantle, as found at high density ($n_{\rm H}\gtrsim 10^4$~cm$^{-3}$) \citep{Bringa04,Wakelam21}.  

\subsection{Desorption rates}

The external UV photodesorption rate can be written as
\begin{equation}
    \Gamma_{ISRF} = I_{ISRF}e^{-\gamma A_\mathrm{v}}Y_{pd},
\end{equation}
where $I_{ISRF}$ is the external interstellar far-UV radiation field (averaged between 6 and 13.6~eV) of $\sim 10^8$~photons~cm$^{-2}$~s$^{-1}$ \citep{Habing1968}; $A_\mathrm{v}$ is the visual extinction; $\gamma$ is a measure of UV extinction relative to visual extinction, which is about 2 for interstellar grains \citep{Roberge1991}; and $Y_{pd}$ is the photodesorption yield of \hhco (in molecules~photon$^{-1}$). The last value was recently determined in the laboratory by \cite{Feraud19},
who were able to derive an intact \hhco photodesorption yield of $4-10\times 10^{-4}$ molecule/photon in various astrophysical environments, including interstellar clouds and prestellar cores (see their Table~1). 
In the following we  use an average experimental yield of $Y_{pd}=7\times 10^{-4}$ molecule/photon for the photodesorption of intact H$_2$CO.

The cosmic-ray induced UV photodesorption rate can be written as
\begin{equation}
    \Gamma_{CRI-UV} = I_{CRI}Y_{pd},
\end{equation}
where $I_{CRI}$ is now the internal UV field induced by the cosmic-ray excitation of H$_2$, which was taken as $I_{CRI}\sim 10^3\zeta_{17}$, where $\zeta_{17}$ is the rate of cosmic-ray ionisation of molecular hydrogen $\zeta$ in units of $10^{-17}$s$^{-1}$ (see the definition of $\zeta$ below). This simple relation between $I_{CRI}$ and $\zeta_{17}$ was deduced from Fig.\,4 of \cite{Shen2004}, with $\zeta$ computed using Eqs.~\ref{flux}-\ref{zeta} below.

Finally, the direct cosmic-ray sputtering rate $\Gamma_{CRD}$ is computed by summing and integrating the product of the sputtering yield $Y_s(\epsilon,Z)$ with the differential cosmic-ray flux $j(\epsilon,Z)$:
\begin{equation}
    \Gamma_{CRD} = 4\pi\sum_Z\int_{0}^{\infty}2Y_s(\epsilon,Z)j(\epsilon,Z)d\epsilon.
  \label{gamma}
\end{equation}
Here $\epsilon$ is the kinetic energy per nucleon, $Z$ is the atomic number of the cosmic-ray nuclei, and $Y_s(\epsilon,Z)$ is the sputtering yield in H$_2$CO/ion obtained by combining the experimental yield $Y_{intact}(S_e)$ for intact sputtering with the calculated electronic stopping power $S_e(\epsilon,Z)$. It is multiplied by a factor of 2 to account for the entrance and exit points of the cosmic-rays through the grain. Assuming a quadratic dependence of $Y_{intact}(S_e)$ with the stopping power, as in Eq. \ref{eq:Ye},
the intact sputtering yield of \hhco molecules can be described as
\begin{equation}
   Y_s(\epsilon,Z) =  Y_{intact}(S_e) = Y_0 S_e^2 = 3.1\times 10^{-4} S_e^2,
\end{equation}
where we combined $Y_{intact} = 2.5\times 10^3$\,molecules\,ion$^{-1}$ (see Table~\ref{Species sputtering yield}) with the stopping power of 2830 eV\,(10$^{15}$\,molecules\,cm$^{-2}$)$^{-1}$ for the $^{86}$Kr$^{18+}$ beam. 

In order to compute $Y_s(\epsilon,Z)$ for a range of kinetic energies (see below), the \texttt{SRIM-2013} code was used to compute the electronic stopping powers $S_e$ for an \hhco ice density of 0.81~g~cm$^{-3}$ and for the cosmic elements with the largest contributions, accounting for the $\sim Z^4$ scaling of the sputtering yields \citep{Dartois23}. In practice, the 15 ions of H, He, C, O, Ne, Mg, Si, S, Ca, Ti, V, Cr, Mn, Fe, and Ni were included using the fractional abundances $f(Z)$ of \citet[their Table\,1]{Kalvans16}. It should be noted that iron, despite its low fractional abundance, gives the largest contribution ($\sim 44$\%) to the rate, as already found  by \citet{seperuelo10} for CO ice.

For the differential cosmic-ray flux $j(\epsilon,Z)$ (in particles~cm$^{-2}$~s$^{-1}$~sr$^{-1}$~(MeV/amu)$^{-1}$) we adopted the functional form proposed by \cite{webber83}
\begin{equation}
  j(\epsilon,Z)=\frac{C(Z)E^{0.3}}{(E+E_0)^3},
  \label{flux}
\end{equation}
where $C(Z)=9.42\times 10^4 f(Z)$ is a normalising constant and $E_0$ a form parameter that varies between 0 and 940~MeV. This formulation corresponds to a primary spectrum since it neglects the energy-loss of cosmic rays through the dense interstellar gas. It allows, however, the full range
of measured ionisation rates to be explored, from standard diffuse clouds (a few $10^{-16}$~s$^{-1}$) to dense clouds (a few $10^{-17}$~s$^{-1}$), by simply varying the parameter $E_0$. In order to infer the relation between $\Gamma_{CRD}$ and $\zeta$, the ionisation rate $\zeta$ was computed as in \citet[their Appendix\,A]{Faure19}:
\begin{equation}
  \zeta=4\pi(1+\beta)(1+\Phi)\int_{0}^{\infty}\sigma_{ion}^{\rm
    p}(\epsilon)j(\epsilon,1)d\epsilon.  
    \label{zeta}
  \end{equation}
Here $\sigma_{ion}^{\rm p}(\epsilon)$ is the cross-section for ionisation of H$_2$ by proton impact, $\Phi$ is a correction factor accounting for the contribution of secondary electrons to ionisation, and where
\begin{equation}
  \beta=\sum_{Z\geq 2}f(Z)Z^2
  \end{equation}
is the correction factor for heavy nuclei ionisation.  Full details and references can be found in Appendix\,A of \citet{Faure19}. In practice, the $E_0$ parameter in Eq.~(\ref{flux}) was taken in the range 200$-$900~MeV so that $\zeta$ was explored in the range $8\times 10^{-18} - 4\times 10^{-16}$\,s$^{-1}$. The integrals in Eqs.~(\ref{gamma}) and (\ref{zeta}) were evaluated numerically from $\epsilon_{\rm min}=100$~eV/amu to $\epsilon_{\rm max}=10$~GeV/amu. The calibration plot between the intact sputtering rate of \ce{H2CO} and the ionisation rate is displayed in Fig.~\ref{fig:calib} (black curve), where $\Gamma_{CRD}$ is found to scale linearly with the ionisation rate. We note that although the low-energy content ($<100$~MeV) of the cosmic-ray spectrum is poorly constrained, the relation between $\Gamma_{CRD}$ and $\zeta$ was found to be quite robust.

\begin{figure}
\includegraphics[angle=0,width=\columnwidth]{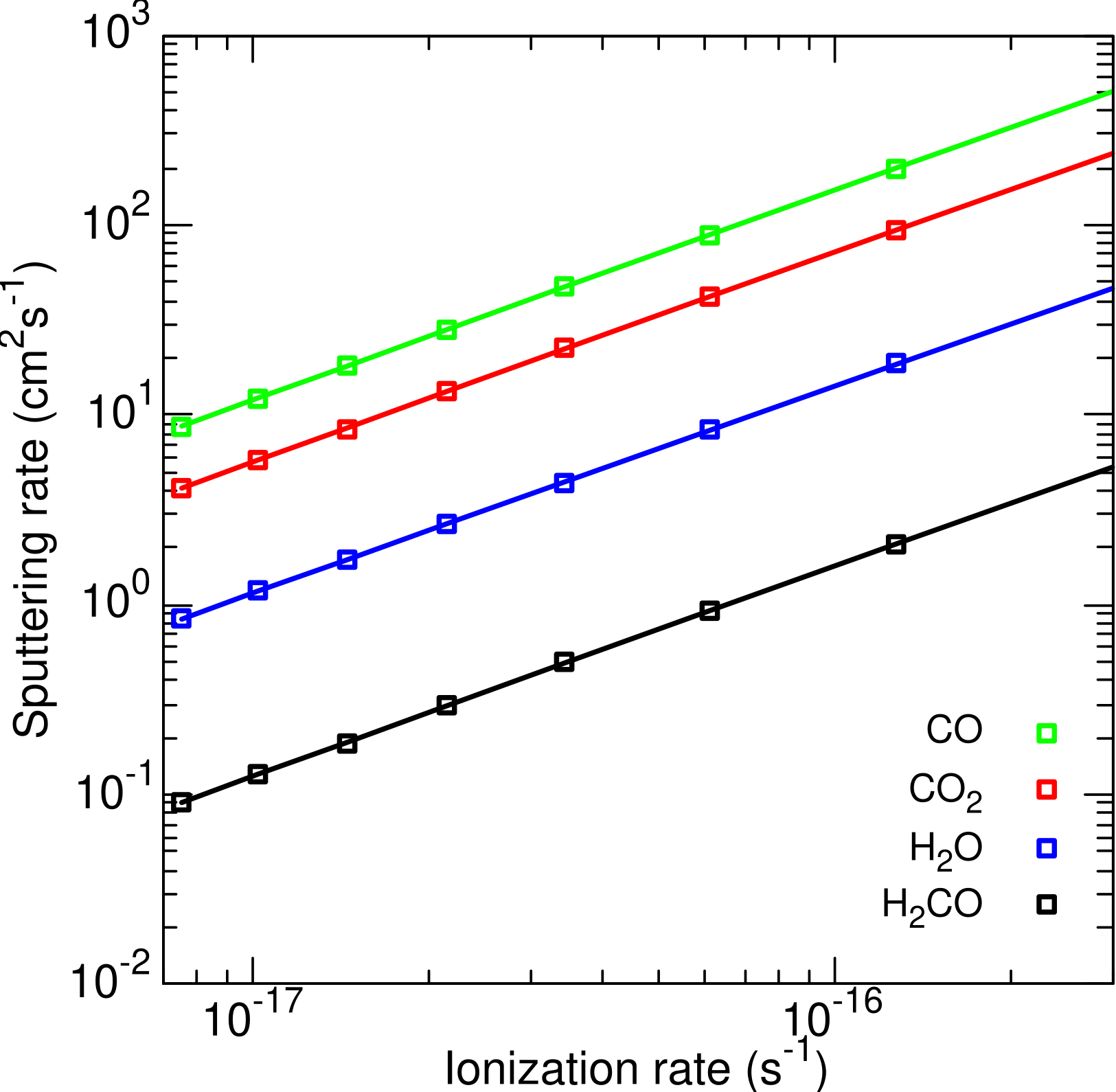}
\caption{Intact sputtering rate of pure \ce{H2CO} and effective sputtering rates of the pure ices \ce{CO}, \ce{CO2}, and \ce{H2O} as a function of the $\zeta$ cosmic-ray ionisation rate. Effective sputtering yield pre-factors are taken from \cite{Dartois23}.}
\label{fig:calib}
\end{figure}

It is now interesting to compare the sputtering rate of \hhco\ with those of the main ice components of interstellar grains, that is H$_2$O, CO, and CO$_2$, where the   latter two have a typical abundance of 20-30\% with respect to H$_2$O. Trace ice species such as NH$_3$, CH$_4$, H$_2$CO, and  CH$_3$OH are expected to be embedded in H$_2$O-rich, CO-rich, or CO$_2$-rich layers. We  assume that the intact sputtering yield of the trapped molecule is equal to the effective sputtering yield of the ice matrix. Simplistic but intuitive, this hypothesis is actually supported by irradiation experiments performed on methanol diluted in H$_2$O and CO$_2$ ice matrices, although the intact sputtering efficiency depends on the dilution factor\footnote{Because \ce{H2CO} easily polymerises, segregation effects could also play a role and reduce the intact sputtering efficiency.
} \citep[see the detailed discussions in][]{Dartois19,Dartois20}. We note that the effective sputtering yield of either water or CO$_2$ was similarly employed to model the cosmic-ray induced sputtering of \ce{CH3OH} in cold molecular cores \citep{Wakelam21, Paulive22}.
The above procedure was thus applied to pure CO, CO$_2$, and H$_2$O ices using the effective sputtering yield pre-factors reported in Table~1 of \citet{Dartois23} with ice densities (needed to compute $S_e$) of 0.88~g~cm$^{-3}$, 1.0~g~cm$^{-3}$, and 0.94~g~cm$^{-3}$, respectively \citep{Luna22,Satorre08,Bouilloud15}. The corresponding sputtering rates as functions of $\zeta$ are shown in Fig.~\ref{fig:calib}. As expected, our results for CO, CO$_2$, and H$_2$O are in very good agreement with those of \citet[see their Fig.\,5]{Dartois23}. We can also observe that the intact sputtering rate of pure \hhco is a factor of 10-100 lower than the effective sputtering rates of H$_2$O, CO, and CO$_2$. 
For convenient use in astrochemical models, simple linear fits of the sputtering rates shown in Fig.~\ref{fig:calib} were performed over the range $\zeta = [8 \times 10^{-18} - 10^{-16}]$~s$^{-1}$. These fits are reported in Table~\ref{tab:yields}.

\begin{table}
\caption{\label{tab:yields}
Fits of direct cosmic-ray sputtering rates $\Gamma_{CRD}$ of pure H$_2$CO, H$_2$O, CO$_2$, and CO ices, as plotted in Fig.~\ref{fig:calib}. See text for details. }
\begin{tabular}{lc}
\hline\hline
 ice   & $\Gamma_{CRD}$ (molecules~cm$^{-2}$s$^{-1}$) \\ 
 \hline
\hhco & 0.15\;$\zeta_{17}\tablefootmark{a}$   \\ 
H$_2$O & 1.2\;$\zeta_{17}$ \\
CO$_2$ & 6.1\;$\zeta_{17}$ \\
CO     & 13\;$\zeta_{17}$ \\     \hline
    \end{tabular}
    \tablefoot{
    \tablefoottext{a}{$\zeta_{17}$ is the rate of cosmic-ray ionisation of molecular hydrogen in units of $10^{-17}$s$^{-1}$. }
    }
\end{table}

\subsection{\ce{H2CO} gas-phase abundance}

Because the total abundance of formaldehyde $n^{tot}_g$ is conserved, we have
\begin{equation}
    n^g_f + n^s_f = n^{tot}_f.
    \label{cons}
\end{equation}
By combining Eqs.~(\ref{ss}-\ref{rdes}) and (\ref{cons}), we obtain
\begin{equation}
    n^g_f = \frac{n^{tot}_f}{1+R_{acc}/R_{des}}=\frac{n^{tot}_f}{1+v_{th}X_{\rm ices}n_{\rm H}/\Gamma^{tot}}
.\end{equation}
Since in practice $v_{th}X_{\rm ices}n_{\rm H}/\Gamma^{tot} \gg 1$ and substituting $v_{th}=\sqrt{8k_BT/(\pi\mu)}$, where $\mu$ is the mass of H$_2$CO, the \hhco gas-phase fractional abundance, $X^g_f = n^g_f/n_{\rm H}$, takes the simple form 
\begin{equation}
    X^g_f = \frac{X^{tot}_f\Gamma^{tot}}{\sqrt{8k_BT/(\pi\mu)}X_{\rm ices}n_{\rm H}},
\end{equation}
where $X^{tot}_f=n^{tot}_f/n_{\rm H}$ is the \hhco total fractional abundance, taken as $3\times 10^{-6}$ (see above). The total fractional abundance of ices, $X_{\rm ices}$, is unknown in L1689B, but again we can assume typical ice abundances in LYSOS, as compiled by \citet[their Table~2]{2015ARA&A..53..541B}. By including `securely', `likely', and `possibly' identified species, we obtain $X_{\rm ices}\sim 8\times 10^{-5}$. Finally, using the physical conditions derived from \hhco emission in L1689B (i.e. $T=10$~K and $n_{\rm H}=4\times 10^5$~cm$^{-3}$), we obtain numerically: 
\begin{equation}
    X^g_f = n^g_f/n_{\rm H} = 1.1\times 10^{-11}\Gamma^{tot}.
    \label{abun}
\end{equation}
In order to reproduce the measured abundance in L1689B ($\sim 6\times 10^{-10}$), a total desorption rate $\Gamma^{tot}\sim 50$~molecules~cm$^2$s$^{-1}$ is required. We compare in Fig.~\ref{fig:model} the observational abundance to our predictions from the above analytical model, assuming the cosmic-ray ionisation rate is $\zeta=3\times 10^{-17}$~s$^{-1}$ (i.e. $\zeta_{17}=3$), which is the typical ionisation rate in molecular clouds, where $A_\mathrm{v}>10$ \citep[see Fig.\,19 in][]{Indriolo12}. Five different models are presented:  desorption is either  due to external and internal photons only (in orange) or  to both photons and direct cosmic-ray sputtering, assuming \hhco ices are pure (in black) or embedded in H$_2$O-rich (in blue), CO$_2$-rich (in red), or CO-rich (in green) matrices. We can observe that photodesorption alone cannot explain the gas phase abundance of \hhco in L1689B and that direct cosmic-ray sputtering is much more efficient, except in the case of a pure \hhco ice. In particular, cosmic-ray  sputtering of \hhco from CO-rich or CO$_2$-rich ices can reproduce the gas phase abundance measured towards L1689B to within a factor of 3. Interestingly, this result is consistent with observational evidence that catastrophic CO freeze-out occurs at $A_\mathrm{v}\geq 9$ and $n_{\rm H}> 10^5$~cm$^{-3}$, where  CO hydrogenation may form a CH$_3$OH-rich layer including  H$_2$CO and new CO$_2$ molecules \citep{2015ARA&A..53..541B}.  It is also possible that part of the interstellar \hhco forms in the gas phase, for example  via the barrierless reaction CH$_3$ + O \citep[see e.g.][]{Terwisscha21}. Thus, while the present results emphasise the potential role of direct cosmic-ray sputtering of \hhco molecules from CO-rich or CO$_2$-rich ices, the importance of gas-phase formation and destruction of \ce{H2CO} will need to be explored in future works. Finally, we note that the \ce{H2CO} gas-phase abundance in L1689B corresponds to about 0.02\% of the total (gas and ice) \ce{H2CO} abundance. 

In summary, our simple model shows that the yields for experimentally determined direct cosmic-ray sputtering are large enough to release significant amounts of \ce{H2CO} ice into the gas-phase, provided that \ce{H2CO} is embedded in CO-rich or \ce{CO2}-rich ice mantles and that \ce{H2CO} molecules co-desorb
with the matrix (i.e. with the same yield). Obviously, further experiments with mixed ices (e.g. \ce{H2CO}:CO), as well as more comprehensive (time-dependent and density-dependent) astrochemical models are needed to confirm these results.

\begin{figure}
\includegraphics[angle=0,width=\columnwidth]{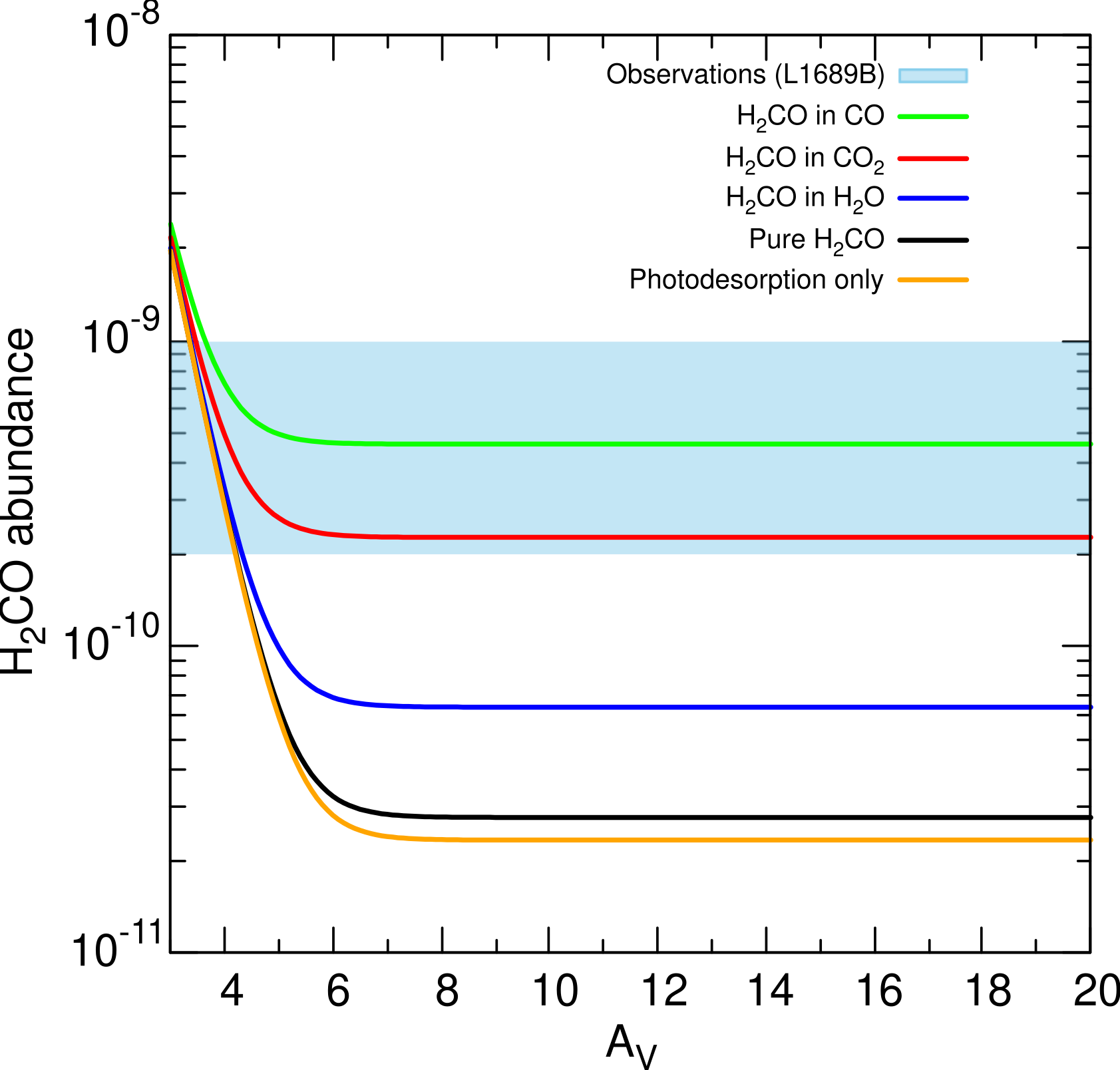}
\caption{\ce{H2CO} gas phase abundance with respect to H nuclei as a function of visual extinction. The observational value derived in this study for the prestellar core L1689B (blue hatched zone) is compared to five different models for \ce{H2CO} sputtering: pure \ce{H2CO} ice and \ce{H2CO} ideally diluted in \ce{H2O}, \ce{CO2}, and CO matrices. See text for details.}
\label{fig:model}
\end{figure}

\section{Conclusions}\label{section:conclusions}

We have studied the sputtering of pure formaldehyde ices under cosmic rays using swift-ion irradiation experiments of H$_2$CO ice films at 10\,K carried out at GANIL with the IGLIAS setup. Formaldehyde is found to quickly polymerise upon swift ion exposure. Other detected radiolysis products in the ice are CO and CO$_2$. We determined the intact H$_2$CO sputtering yield in the gas phase to be $Y_\mathrm{intact}= 2.5\times 10^3$ molecule\,ion$^{-1}$ for a stopping power of 2830 eV/10$^{15}$ molecule\,cm$^{-2}$. Assuming a quadratic dependence of the sputtering yield with stopping power, this corresponds to an intact sputtering prefactor $Y_0 =  3.1\times 10^{-4}$ $10^{30}$~eV$^{-2}$cm$^{-4}$molecule$^3$. The effective sputtering prefactor is $Y_0^{eff}= 1.4\times 10^{-3}$ $10^{30}$~eV$^{-2}$cm$^{-4}$molecule$^3$.
With these values we determined that the direct cosmic-ray sputtering rate for intact H$_2$CO in pure ices can be expressed  simply by $\Gamma_\mathrm{CRD}(\mathrm{H_2CO}) = 0.15\,\zeta_{17}$ molecules\,cm$^{-2}$\,s$^{-1}$, where $\zeta_{17}$ is the cosmic-ray ionisation rate in units of 10$^{-17}$ s$^{-1}$. 

Using a simple steady-state chemical model including accretion of molecules from the gas phase onto the grains and their desorption by either UV photons (both external and secondary UV photons produced by cosmic-ray excitation of H$_2$) or direct cosmic-ray impact, we find that photodesorption is not efficient enough to account for the gas-phase abundance of H$_2$CO of  $6\times 10^{-10}$ (with respect to H nuclei) measured in the prestellar core L1689B, nor  is cosmic-ray induced sputtering of a pure H$_2$CO ice. 
On the other hand, when H$_2$CO is diluted in CO or in CO$_2$ ices, our model can reproduce the H$_2$CO gas-phase abundance, assuming that minor ice constituents like formaldehyde will be sputtered to the gas phase at the same effective rate as the ice mantle bulk. However, the cosmic-ray  sputtering rate is  insufficient if H$_2$CO is embedded in a water ice mantle. Future experimental studies should be dedicated to investigating the polymerisation and sputtering of \hhco diluted in other ice matrices (e.g. H$_2$O, CO, CO$_2$). The competition between the cosmic-ray induced sputtering and the gas-phase chemistry of \ce{H2CO} will also need to be explored.

\begin{acknowledgements}
      This work is based on observations carried out under project numbers 090-17 and 135-18 with the IRAM 30m telescope. IRAM is supported by INSU/CNRS (France), MPG (Germany) and IGN (Spain). This research has made use of spectroscopic and collisional data from the EMAA database (https://emaa.osug.fr and https://dx.doi.org/10.17178/EMAA). EMAA is supported by the Observatoire des Sciences de l’Univers de Grenoble (OSUG).
      The experiments were performed at the Grand Accélérateur National d’Ions Lourds (GANIL) by means of the CIRIL Interdisciplinary Platform, part of CIMAP laboratory, Caen, France. We thank the staff of CIMAP-CIRIL and GANIL for their invaluable support. We acknowledge funding from the National Research Agency, ANR IGLIAS (grant ANR-13-BS05-0004). Thibault Launois is acknowledged for his contribution to the experiments and data reduction.
\end{acknowledgements}

%
%
\bibliographystyle{aa}
\bibliography{biblio}

%
%
%
%
%
%
%



\begin{appendix}

 \section{Comparison of the experimental and interstellar doses}\label{appendix:dose}

In our experiments, the dose $D$ ($D=S_e \cdot F$ where $F$ is the fluence) lies between 0.0025 and 12 eV/molecule at the investigated experimental fluences ($8.8 \times 10^{8}-4.4\times 10^{12}$\,ion\,cm$^{-2}$). These values can be compared to the "interstellar" dose received by a grain over 1\,Myr (the typical age of a molecular cloud or prestellar core), which we estimate as $\sim 0.3$~eV/molecule using the cosmic-ray flux given in Eq.~\ref{flux}, with $E_0=500$~MeV and $S_e$ computed with the \texttt{SRIM-2013} code. As a result, the most astrophysically relevant conditions in our experiment should be the early times of the measurements, when the fluence is lower than a few $\sim 10^{11}$~ions\,cm$^{-2}$. We note, however, that the experimental and calculated doses correspond to different kinds of ions, Kr and essentially H, He, C, O, and Fe, respectively. It is unclear whether the nature of ions has an effect on radiolysis. Thus, threshold effects have been reported for polymers that contain aromatic species, but not in the case of aliphatic polymers that are chemically more similar to ices \citep{1995NIMPB.105...46B}. To our knowledge, no similar studies have been published for ices, and firm conclusions about the relevance of experimental doses would require comparative radiolysis investigations with e.g. H, He and Fe ions.

\section{H$_2$CO spectra}\label{appendix:spectra}

The spectra observed with the IRAM 30\,m radiotelescope and used to determine the H$_2$CO column density are presented in Fig.\,\ref{fig:h213co_spectra} for H$_2^{13}$CO and in Fig.\,\ref{fig:h2c18o_spectra} for H$_2$C$^{18}$O. In the figures,`o' and `p' stand for ortho and para spin symmetry, respectively. The spectra were taken at coordinates RA = 16$^h$34$^m$48.3$^s$, Dec = $-24^\circ 38^\prime 04^{\prime\prime}$.
\begin{figure*}
	\includegraphics[angle=0,width=17cm]{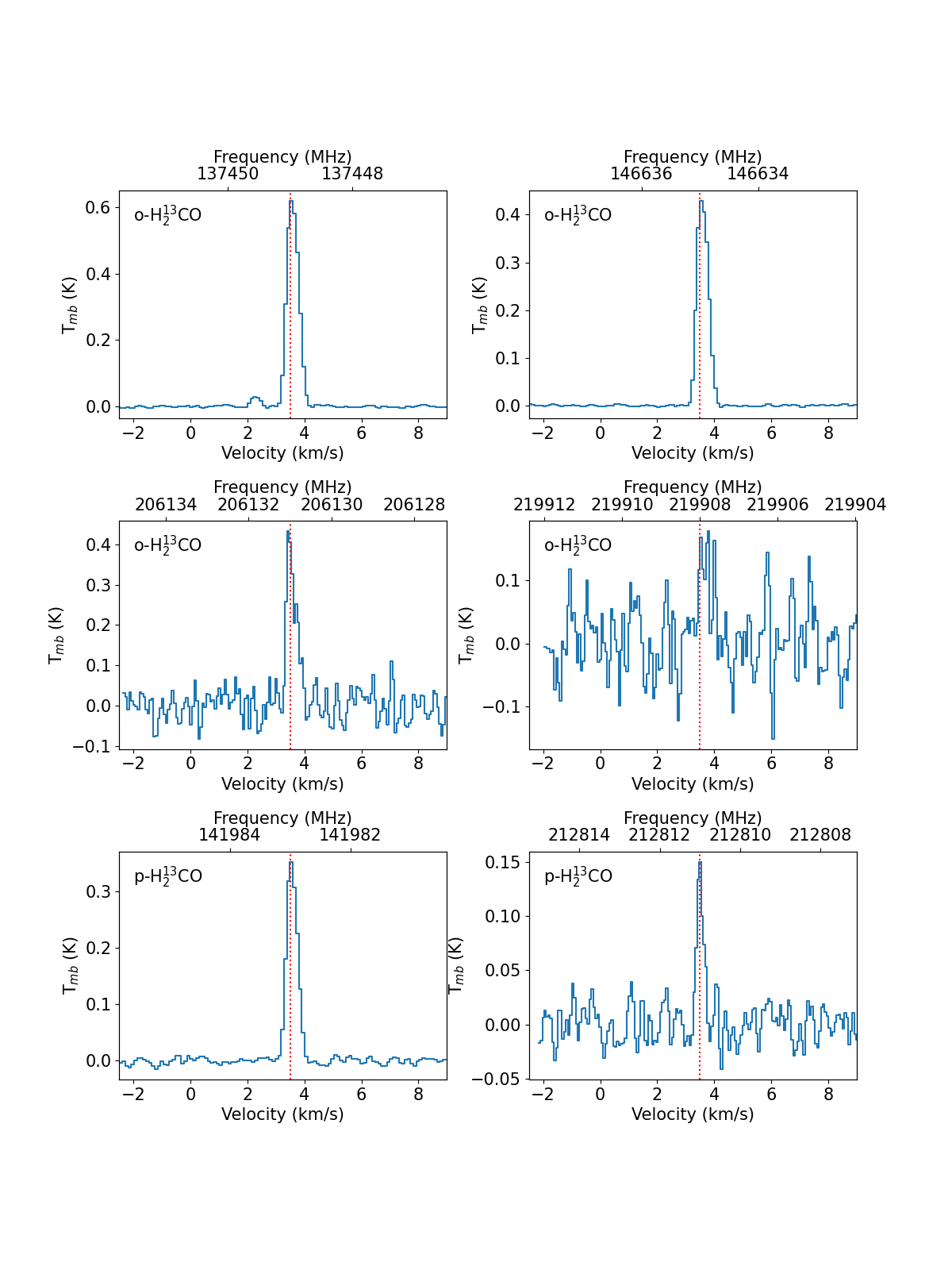}
	\caption{Observed H$_2^{13}$CO spectra. The vertical dotted line shows the source velocity ($v=3.5$\,km\,s$^{-1}$). The spin symmetry is indicated in the upper left corner of each plot, with `o' standing for ortho and `p' standing for para.}
	\label{fig:h213co_spectra}
\end{figure*}

\begin{figure*}
	\includegraphics[angle=0,width=17cm]{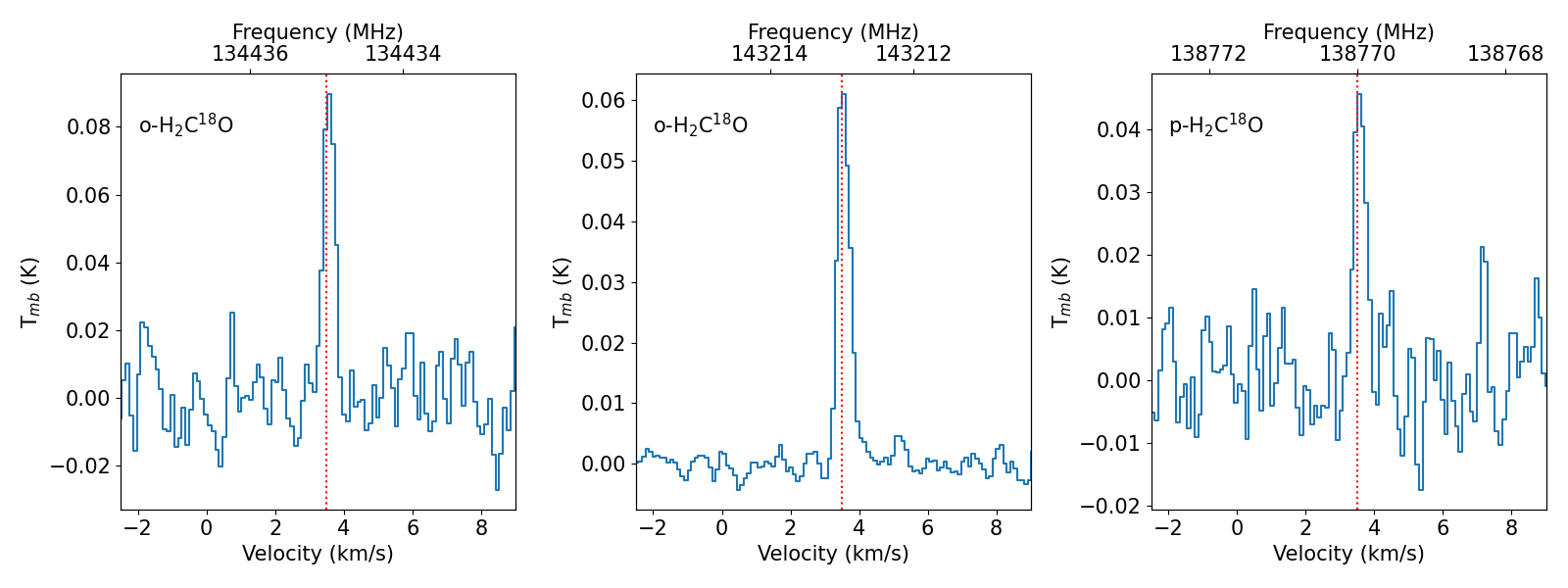}
	\caption{Observed H$_2$C$^{18}$O spectra. The vertical dotted line shows the source velocity ($v=3.5$\,km\,s$^{-1}$).The spin symmetry is indicated in the upper left corner of each plot, with `o' standing for ortho and `p' standing for para.}
	\label{fig:h2c18o_spectra}
\end{figure*}

\FloatBarrier

\section{Sputtering yield determination: The case of a thin CO$_{2}$  ice film}\label{appendix:co2}

High-energy heavy ions irradiation of \ce{CO2} ices has been thoroughly investigated \citep{seperuelo09,Mejia15,Dartois21}. In this appendix, we compare sputtering yields estimated from the interference fringes analysis method (Method 1, as employed in this work for H$_2$CO ice) to yields derived from the integration of IR bands (Method 2).

\subsection{Method 1: Interference fringes method}

The procedure follows the steps explained in Section\,\ref{section:experimentalresults} for H$_2$CO ice. The interference fringes observed in the IR spectra during irradiation are depicted in Fig.~\ref{fig:CO2_Transmittance}. The adjustments were achieved using Eq.~\ref{eq:interference}. While a more rigorous equation including the optical indices could be applied in the case of \ce{CO2} \citep[see][for details]{Dartois23}, we opted for the simplified equation (Eq. \ref{eq:interference}) in order to be consistent with the methodology employed in this study. The total mass sputtering yield ($Y_{m,tot} = (5.7\pm0.1) \times 10^{-18}$ g\,ion$^{-1}$) was extracted from the linear fit of surface density as a function of fluence (Fig.\,\ref{fig:CO2_MassPerArea}), using a density of CO$_2$ of 1.0\,g\,cm$^{-3}$ \citep{Bouilloud15}. We note that in the case of CO$_2$, there is no visible compaction phase, despite experimental conditions close to those of H$_2$CO (similar film thicknesses, pressure and temperature, see Section\,\ref{section_experimental}).

\begin{figure}
\includegraphics[angle=0,width=0.95\columnwidth]{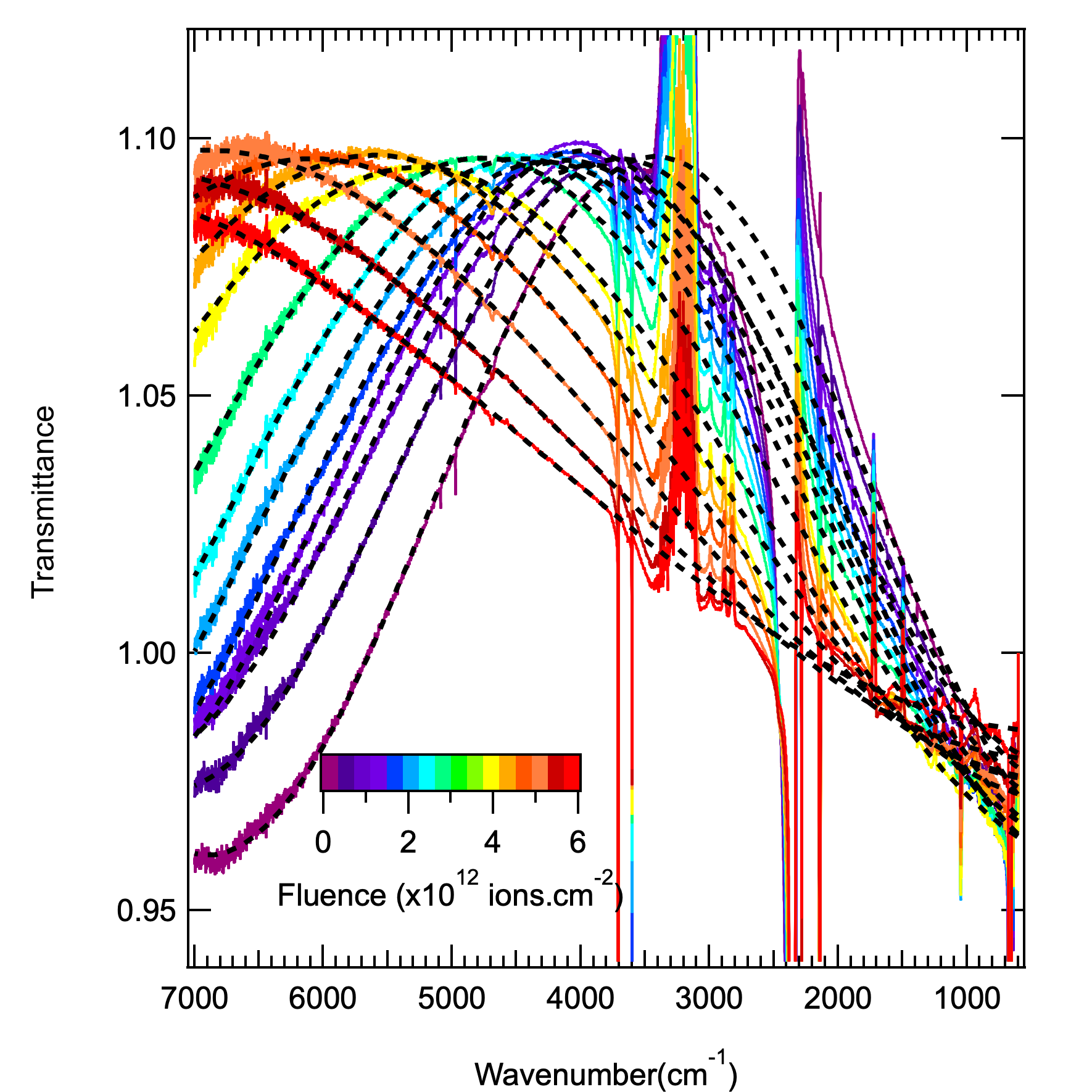}
\caption{Infrared transmittance spectra of a CO$_{2}$ ice film evolution upon 74 MeV $^{86}$Kr$^{18+}$ swift heavy ion irradiation. \emph{Solid lines}: Experimental data. \emph{Dashed lines}: Model spectra fitted to the data, without considering the absorption bands. The QMS signal ends around a fluence of $4.10^{12}$ ion\,cm$^{-2}$, before the end of the irradiation as the instrument stopped prematurely.}
\label{fig:CO2_Transmittance}
\end{figure}

\begin{figure}
\includegraphics[angle=0,width=0.95\columnwidth]{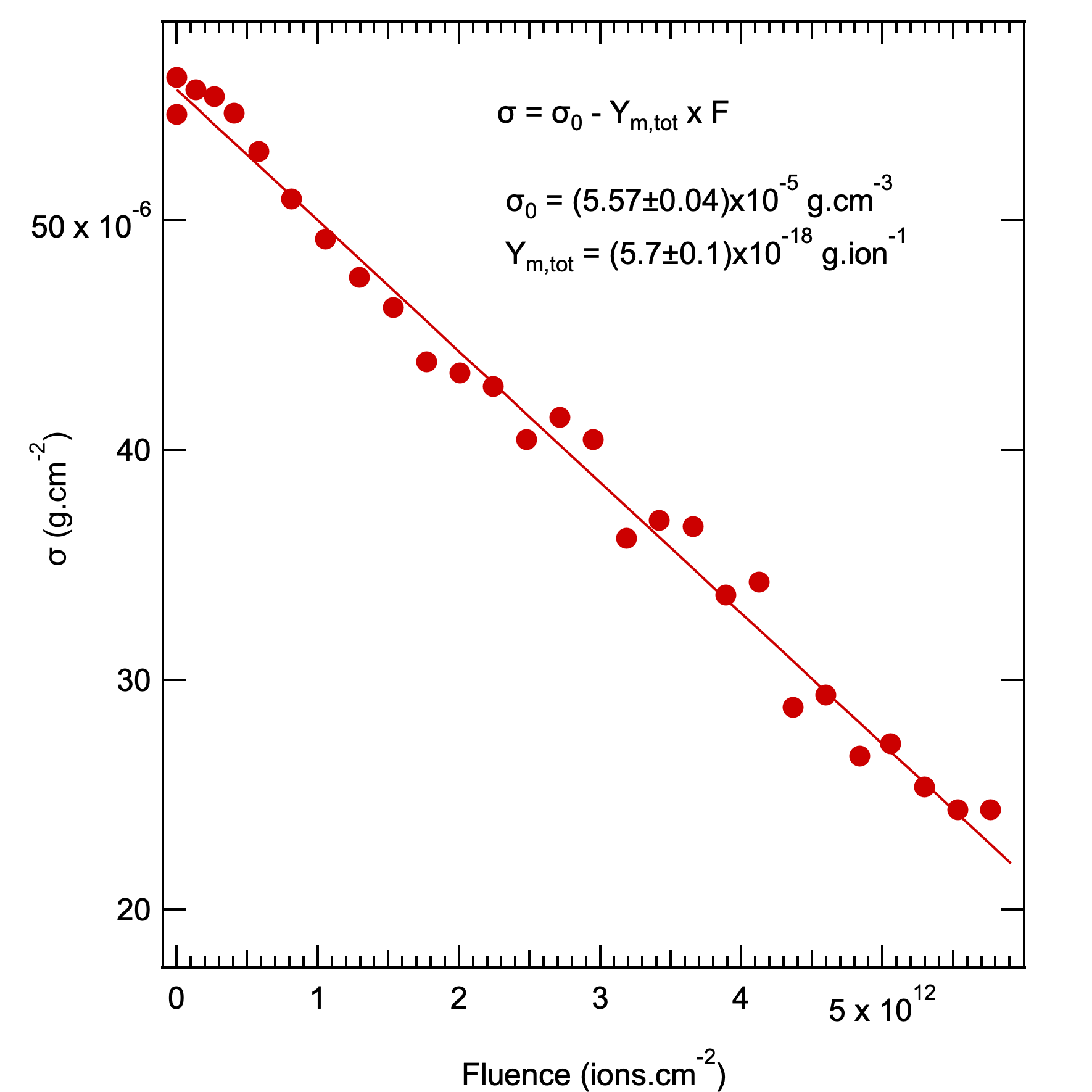}
\caption{Evolution of the bulk mass per area as a function of fluence deduced from interference fringes for a CO$_2$ ice film.}
\label{fig:CO2_MassPerArea}
\end{figure}

The CO$_2$ fragmentation pattern was derived from the QMS signal during CO$_2$ injection. It is composed of only four fragments at $m/z$=44, $m/z$=28, $m/z$=16, and $m/z$=12 (with intensity ratios of 0.748, 0.112, 0.07, and 0.07, respectively). Figure~\ref{fig:CO2_QMS} shows the molecular fraction deduced from the QMS signal after (i) correction of the background, (ii) extracting the original CO$_2$ signal intensity using the fragmentation pattern ($I_{CO_{2}}=I(m/z=44)/0.748$), (iii) subtracting the CO$_2$ fragments from other mass signals (e.g. $I_C=I(m/z=12)-0.07 \times I_{CO_{2}}$), and (iv) dividing by the total electron-impact ionisation cross sections ($\sigma^{impact}$(CO$_2$)=3.521$\AA^2$, from the NIST database ).

In contrast to H$_2$CO, no significant variation of the \ce{CO2} fraction is observed during the very early fluences; however, a similar smooth and continuous evolution is observed thereafter. Over the entire experiment, the molecular fraction of CO increases by $\sim 40\%$, while that of \ce{CO2} decreases by $\sim60\%$. The sputtering yields were calculated using Eq. \ref{eq:fm} after fitting the molecular fractions with a third-order polynomial, as in the case of \ce{H2CO}. The fits are shown as dashed lines in the bottom panel of Fig.~\ref{fig:CO2_QMS}. The average sputtering yields are provided in Table\,\ref{tab:yields}.

\begin{figure}
\includegraphics[angle=0,width=\columnwidth]{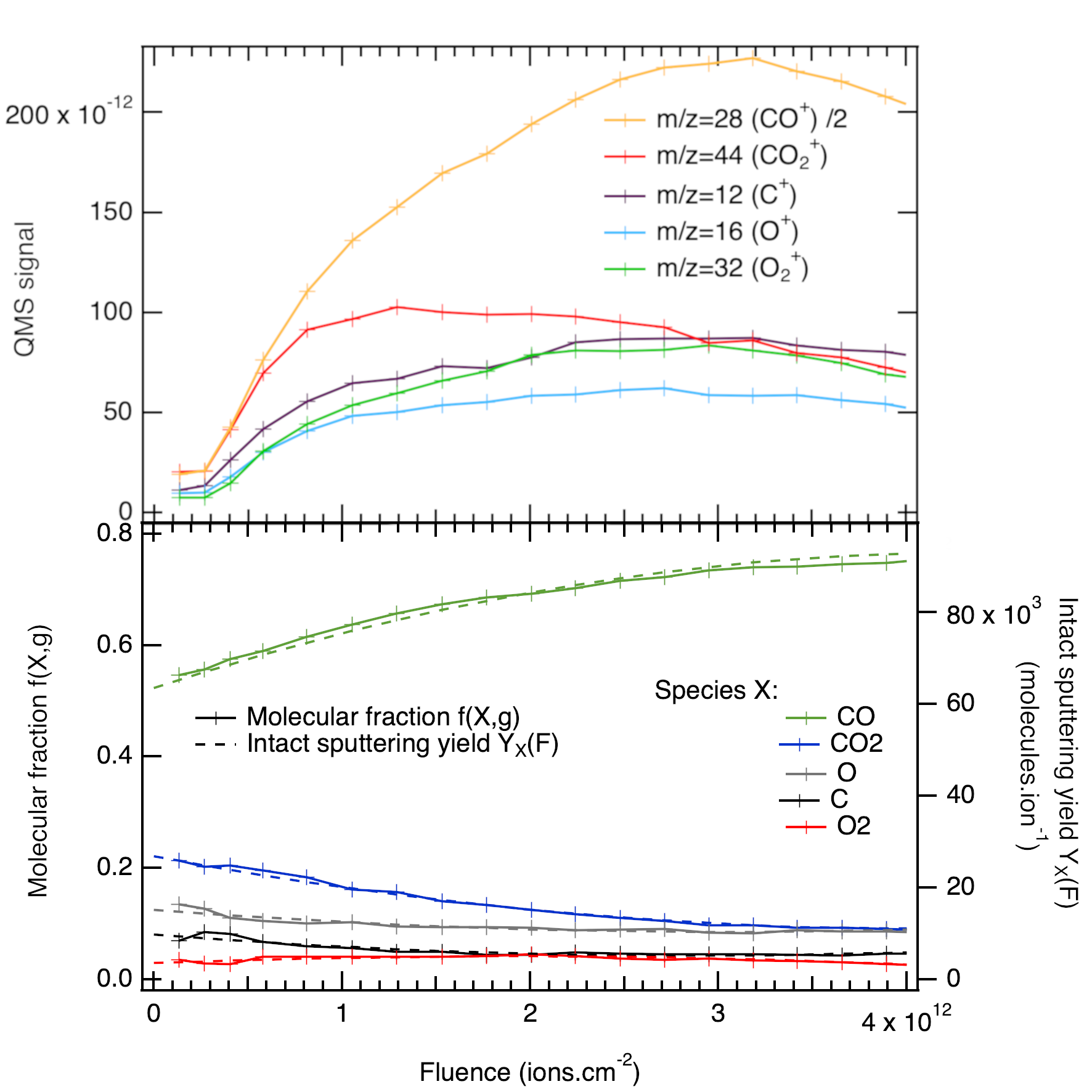}
\caption{QMS signals as a function of fluence. \emph{Top panel}: Background corrected signals. \emph{Bottom panel}: Molecular fractions in the gas (\emph{solid lines}, left $y$-axis) derived from the fragmentation patterns and ionisation cross sections, and sputtering yield of each species (\emph{dashed lines}, right $y$-axis). See text for details.}
\label{fig:CO2_QMS}
\end{figure}

\begin{table}
\caption{\label{CO2 Species sputtering yield} Initial gas phase species fractions $f_{i}(X,g)$ and average sputtering yield $Y_{X}$ for each species $S$.}
\centering
\begin{tabular}{cccc}
        \hline\hline
        Species  &  $f_{i}(X,g)$  &$Y_{X}$  \\
         (X)& & molecules\,ion$^{-1}$\\
        \hline
        CO      &0.55  & $(8.0\pm3.2)\times10^4$ \\
        CO$_{2}$&0.21 & $(1.7\pm1.0)\times10^4$ \\
        O      & 0.14  & $(1.2\pm0.4)\times10^3$\\
        C      & 0.07 & $(6.7\pm2.0)\times10^3$\\
        O$_{2}$ & 0.04 & $(4.4\pm0.9)\times10^3$ \\
        \hline
    \end{tabular}
\end{table}

As in the case of \ce{H2CO}, the intact fraction of sputtered CO$_2$ was computed using the initial fractions \(f_i(X,g)\) of species containing carbon atom (Table \ref{CO2 Species sputtering yield}) :
\begin{equation}
    \eta = \frac{f_i({\rm CO_2},g)}{f_i({\rm CO_2},g)+ f_i({\rm CO})+f_i({\rm C},g)}.
\end{equation}

We obtain $\eta$=0.25$\pm$0.02 from which the effective sputtering yield (see Eq.~\ref{eq:Yeff}) for a CO$_2$ ice film was estimated as $Y_{eff}$=(6.8$\pm$4.1)$\times$10$^{4}$ molecules\,ions$^{-1}$. The corresponding effective prefactor derived from Eq.~\ref{eq:Ye} is $Y_0^{eff}$=(3.8$\pm$2.3)$\times$10$^{-3}$\,10$^{30}$\,eV$^{-2}$\,cm$^{-4}$\,molecule$^{3}$, using a stopping power of 3290~eV/10$^{15}$ molecules\,cm$^{-2}$. 


\subsection{Method 2: Integrated bands method}

Fig.~\ref{fig:CO2_IR_spectra} illustrates the evolution of the CO$_2$ infrared vibrational band at 2343 cm$^{-1}$ with fluence. The integrated intensity of this band ($A$) was employed to derive the CO$_2$ column density ($N$) with the first-order relation:
\begin{equation}
    N = \frac{\ln(10)}{\mathcal{A}} \int A(\nu) \, \mathrm{d}\nu,
\end{equation}
where the band strength ($\mathcal{A}$) is $7.6 \times 10^{-17}$ cm\,molecules$^{-1}$ \citep{Bouilloud15}. The derivative of the column density with respect to fluence was fitted using an equation involving both radiolysis and sputtering processes:
\begin{equation}
    -\frac{dN}{dF} = \sigma_{des}N + Y_{eff}(1 - e^{-\frac{N}{N_{d}}})f,
    \label{dNdF}
\end{equation}
where $\sigma_{des}$ is the destruction cross-section of CO$_2$, $f$ is the relative concentration of carbon dioxide molecules with respect to the total number of molecules in the ice film, and $N_d$ is the semi-infinite sputtering depth. Similarly to the experiment of \cite{Mejia15} on CO$_2$ ice irradiation, the fraction of CO$_2$ in the ice remains significantly higher than that of CO at the observed fluences, and the value of $f$ is set to 1. Adjusting the data with Eq.~\ref{dNdF}, as shown in Fig.~\ref{fig:CO2_mdNdF}, provides the following solution: $Y_{eff} = (7.0 \pm 0.9) \times 10^{4}$ molecules\,ions$^{-1}$, $\sigma_{des} = (1.4 \pm 0.1) \times 10^{-13}$ cm$^{2}$\,ion$^{-1}$, and $N_{d} = (1.8 \pm 0.3) \times 10^{17}$ molecules\,cm$^{-2}$. The derived effective sputtering yield corresponds to an effective prefator $Y_0^{eff}$ = (6.5 $\pm$ 0.8) $\times$ 10$^{-3}$\,10$^{30}$\,eV$^{-2}$\,cm$^{-4}$\,molecule$^{3}$, using again a stopping power of 3290\,eV/10$^{15}$ molecules\,cm$^{-2}$.

\begin{figure}
\includegraphics[angle=0,width=\columnwidth]{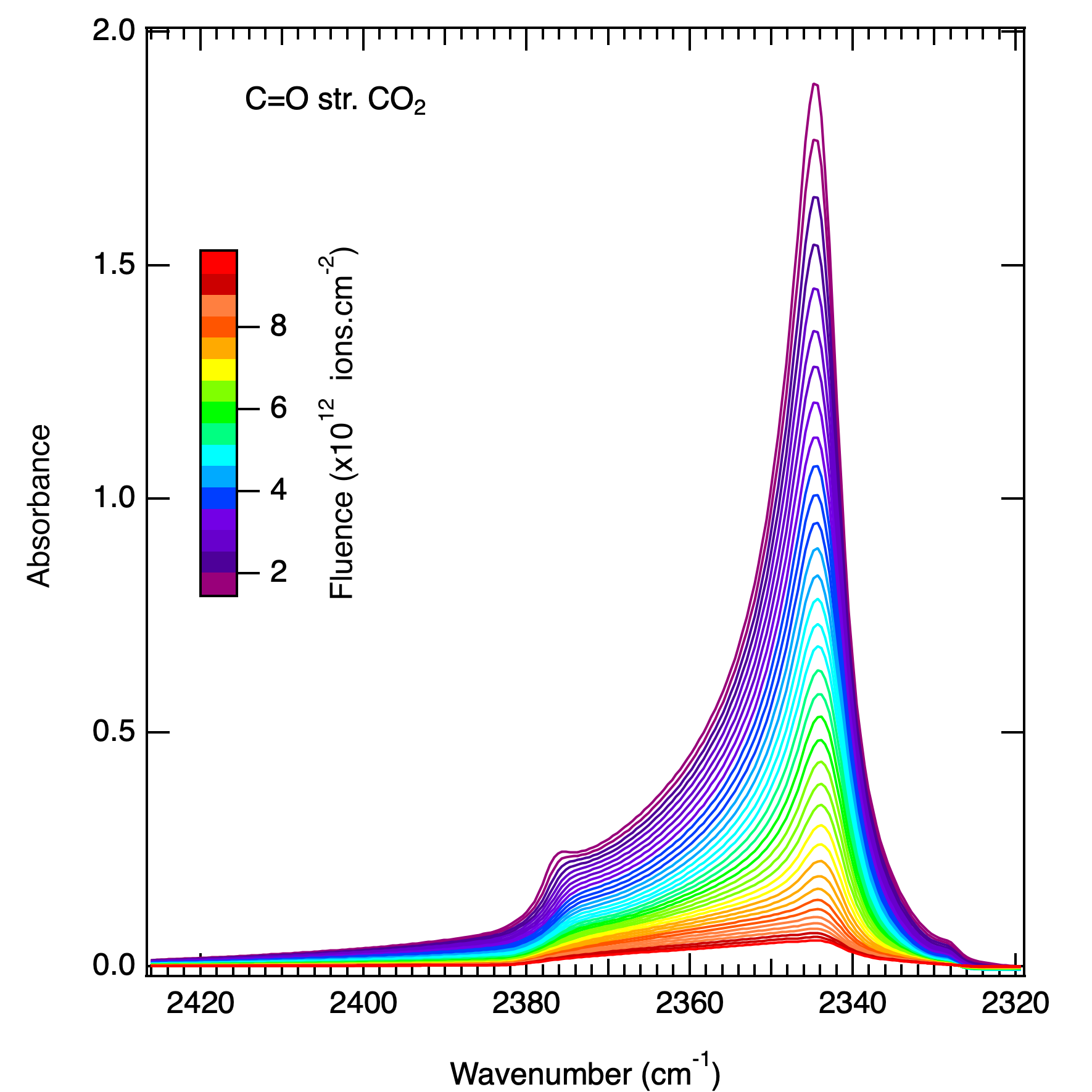}
\caption{Evolution of the background-corrected IR spectrum centered on the $\nu_3$ stretching mode of \ce{CO2} upon 74~MeV $^{86}$Kr$^{18+}$ swift ion irradiation}
\label{fig:CO2_IR_spectra}
\end{figure}

\begin{figure}
\includegraphics[angle=0,width=\columnwidth]{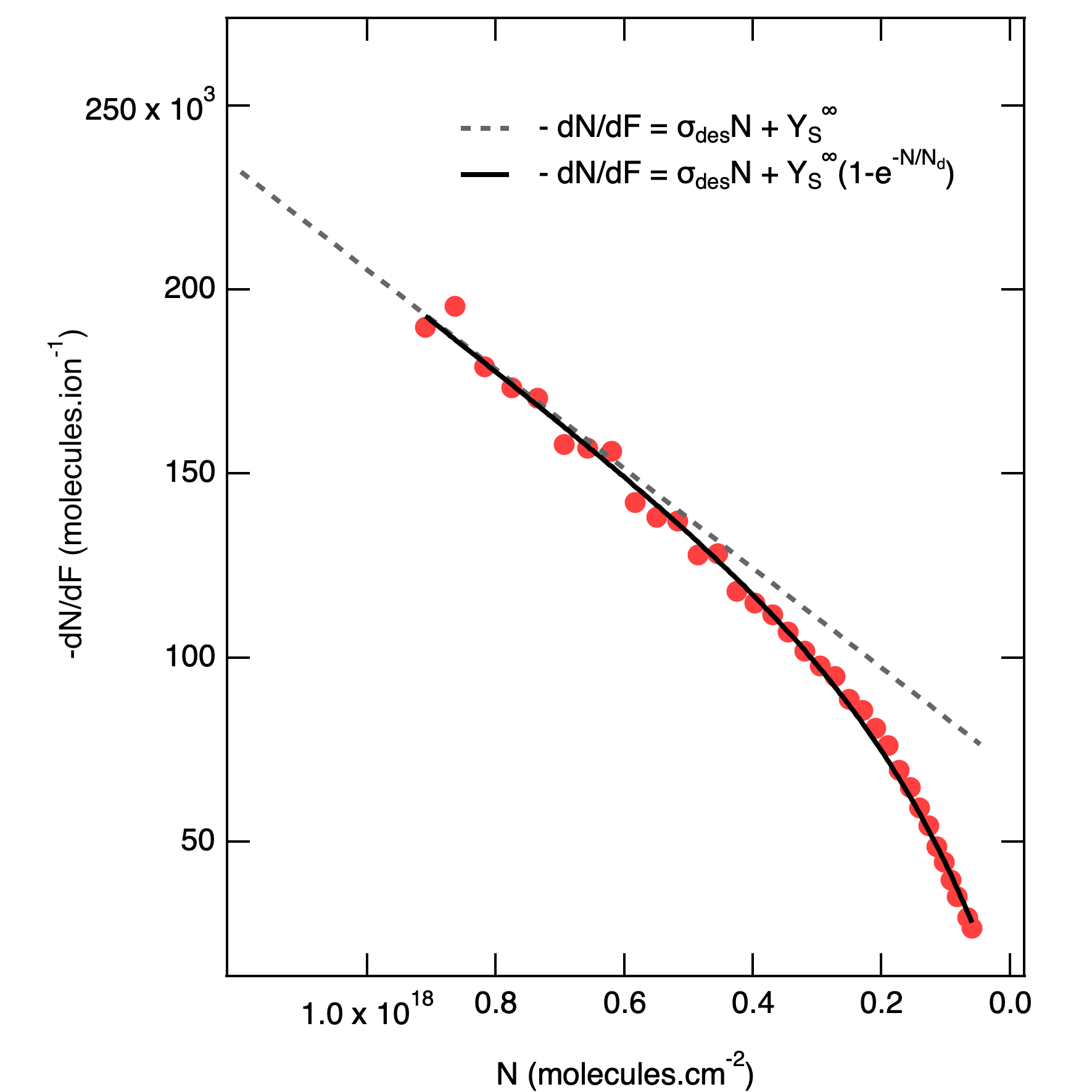}
\caption{Column density derivative with respect to fluence. The solid and dashed lines are fits with or without including the sputtering depth $N_d$ in Eq.~\ref{dNdF}, respectively.}
\label{fig:CO2_mdNdF}
\end{figure}

\subsection{Comparison of Method 1 and 2}

Results from this work and previous studies are presented in Table~\ref{tab:CO2_Yield_comparison}. The comparison with literature data is not straightforward due to various experimental conditions. For example, experiments have shown that the sputtering efficiency depends on the ice thickness: the effective sputtering yields $Y_{\text{eff}}$ for CO and \ce{CO2} ices were indeed found to be constant for thick films (more than $\sim$ 100 monolayers) but to decrease significantly for thin ice films \citep{Dartois21}. We recall that $Y_{\text{eff}}$ scales with the square of the heavy ion stopping power, while $Y_e^0$ should be roughly constant for a given target. We can observe in Table~\ref{tab:CO2_Yield_comparison} that our $Y_e^0$ value derived from the integrated band method (Method 2) agrees with those of \cite{seperuelo09} and \cite{Dartois21}, who used the same method, to better than a factor of 2, and almost within error bars (whatever the thickness of the film). In addition, and more importantly, the interference fringes method (Method 1) gives an effective sputtering yield prefactor ($Y_e^0$ value)  in very good agreement (i.e. within error bars) with the value we derived from Method 2.

We conclude that the interference fringes method, even using a simplified optical model, is a powerful alternative to the integrated bands method for the determination of sputtering yields. Furthermore, combined with QMS measurements, this method allows us to determine both effective and intact sputtering yields, this latter being challenging to estimate with other methods. 

\begin{table*}[ht!]
\caption{\label{tab:CO2_Yield_comparison} Comparison of results for swift heavy ion irradiation of pure CO$_2$ ice films.}
\centering
\begin{tabular}{cccccc}\hline\hline
        Method & Reference & $d_0$\tablefootmark{a}& $S_e$  & $Y_{\text{eff}}$ ($\times$ 10$^{3}$)  & $Y_e^0$ ($\times$10$^{-3}$)\\
               &           & ($\mu$m)&(eV/10$^{15}$ molecules\,cm$^{-2}$)&(molecule\,ions$^{-1}$)\\
        \hline
        Interference fringes & This work & 0.81 &3290 & 68 $\pm$ 41 & 6.3 $\pm $3.4\\
        \hline 
         & This work &0.81 &3290 & $70 \pm 9$ & $6.5 \pm 0.8$ \\
         Column density& \cite{seperuelo09} & 1.08 &2601.9 & $78 \pm 31$ & $11.5 \pm 4.5$ \\
         from band& \cite{Dartois21} & 0.51 &2487.8 & $37.6 \pm 23.8$ & $6.1 \pm 3.8$\\
        integration & \cite{Dartois21} & 1.67 &2487.8 & $48.2 \pm 17.0$ & $7.8 \pm 2.7$\\
         & \cite{Dartois21} &  0.80 &1919.1 & $15.5 \pm 3.6$ & $4.2 \pm 1.0$ \\
         \hline
\end{tabular}
\tablefoot{
\tablefoottext{a}{Initial thickness of the film, before irradiation.}
}
\end{table*}

\end{appendix}
\end{nolinenumbers}
\end{document}